\documentclass[10pt,doublecolumn,journal]{IEEEtran}
\usepackage{latexsym}
\usepackage{amsfonts}
\usepackage{amsbsy}
\usepackage{amsmath,amssymb}
\usepackage{times}
\usepackage{graphicx}
\usepackage{enumerate}
\usepackage[usenames]{color}
\usepackage[dvips]{pstcol}
\usepackage{epstopdf}

\usepackage{subeqnarray}
\usepackage{cases}
\usepackage{stfloats}
\usepackage{cases}
\usepackage{subeqnarray}
\usepackage{amsmath} 
\usepackage{multicol}
\usepackage{multirow}
\usepackage{bigstrut}
\usepackage{amsthm}
\usepackage{booktabs}
\usepackage{bm}
\usepackage{algorithm}
\usepackage{algorithmic}
\usepackage{bm}
\usepackage{amsthm}
\usepackage{cite}
\input epsf

\title{Outage-Constrained Robust Beamforming for Intelligent Reflecting Surface Aided Wireless Communication} 
\author{Ming-Min Zhao, \IEEEmembership{Member,~IEEE,} An Liu, \IEEEmembership{Senior Member,~IEEE,} and Rui Zhang, \IEEEmembership{Fellow,~IEEE}
	\thanks{
		M. M. Zhao and A. Liu are with the College of Information Science and Electronic Engineering, Zhejiang University (email: \{zmmblack, anliu\}@zju.edu.cn). R. Zhang is with the Department of Electrical and Computer Engineering, National University of Singapore (email: elezhang@nus.edu.sg). 
	}
}

\begin{document}
		\maketitle
	\begin{abstract} 
	In intelligent reflecting surface (IRS) aided wireless communication systems, channel state information (CSI) is crucial to achieve its promising passive beamforming gains. However, CSI errors are inevitable in practice and generally correlated over the IRS reflecting elements due to the limited training with discrete phase shifts, which degrade the data transmission rate and reliability. In this paper, we focus on investigating the effect of CSI errors to the outage performance in an IRS-aided multiuser downlink communication system. Specifically, we aim to jointly optimize the active transmit precoding vectors at the access point (AP) and passive discrete phase shifts at the IRS to minimize the AP's transmit power, subject to the constraints on the maximum CSI-error induced outage probability for the users. First, we consider the single-user case and derive the user's outage probability in terms of the mean signal power (MSP) and variance of the received signal at the user. Since there is a trade-off in tuning these two parameters to minimize the outage probability, we propose to  maximize their weighted sum with the optimal weight found by one-dimensional search. Then, for the general multiuser case, since the users' outage probabilities are difficult to obtain in closed-form due to the inter-user interference, we propose a novel constrained stochastic successive convex approximation (CSSCA) algorithm, which replaces the non-convex outage probability constraints with properly designed convex surrogate approximations. Simulation results verify the effectiveness of the proposed robust beamfoming algorithms and show their significant performance improvement over various benchmark schemes.
	\end{abstract}  
	\begin{IEEEkeywords} 
		Intelligent reflecting surface, channel estimation error, robust beamforming, outage probability, phase-shift optimization.
	\end{IEEEkeywords}
	
\section{Introduction}
For the fifth-generation (5G) wireless communication networks that are being standardized and deployed worldwide, various transmission technologies such as massive multiple-input multiple-output (MIMO), ultra-dense network (UDN) and millimeter wave (mmWave) communication have been adopted to meet the ever-increasing requirements in terms of data rate, reliability, latency and connectivity \cite{Boccardi2014}. However, these technologies face similar challenges in practical implementation due to their required high hardware cost and energy consumption. Moreover, they only adapt to the time-varying radio environment to some extent and thus cannot always guarantee the quality-of-service (QoS), especially in harsh propagation environment with severe signal blockage or deep fading. 
Recently, intelligent reflecting surface (IRS) has emerged as a promising technology to enhance the spectral efficiency of wireless communication systems cost-effectively \cite{Wu2019Magazine, Liaskos2018, Renzo2019, Basar2019, Wu2020Tutorial, Huang2020HMIMO}. Specifically, IRS is a planar surface composed of a large number of passive reflecting elements, each of which can induce an independent phase shift and/or amplitude change of the incident signal in real-time. Thus, IRS is able to program/reconfigure the signal propagation by dynamically adjusting its reflection coefficients based on the channel state information (CSI), and achieve cost-effective performance improvement with low hardware and energy cost.

As such, IRS has attracted significant attention recently and its reflection optimization has been investigated in various aspects and under different setups (see, e.g., \cite{Hu2018TSP, Wu2018_journal, Cui2019, Jiang2019, Huang2019, zhao2019intelligent, ZhangMIMO, zuo2020resource, Ding2020_IRSNOMA, Huang2020DRL}), where IRS was shown to be effective in enhancing the system performance substantially. However, the above performance gains in IRS-aided communication systems are crucially dependent on the CSI of the IRS-associated channels, which is practically challenging to obtain due to the following reasons. First, since IRS is generally a passive device and does not have power amplifies, the conventional channel estimation approach relying on the pilot signal sent by the IRS is inapplicable. Second, the number of channel coefficients in IRS-associated channels is enormous due to massive reflecting elements at the IRS, especially when the access point (AP) and/or users are equipped with multiple antennas, which makes accurate channel estimation practically difficult given limited channel training power and time. To address the above challenges, various methods have been proposed to estimate the IRS channels, see, e.g., \cite{He2019_CE, you2019progressive, wang2019channel, chen2019channel, zheng2019intelligent, wei2020channel} and the references therein. However, despite the rapid progress in channel estimation studies for IRS-aided systems, channel estimation errors are practically inevitable and their impact on the system performance needs to be taken into account for the IRS-aided data transmission.

In the literature, there have been some recent works that studied robust beamforming designs for IRS-aided communication systems under imperfect CSI \cite{you2019progressive, yu2019robust, Zhou2020robust, guo2019weighted, Zhao2020intelligent, zhou2020framework}, which are generally based on three different design approaches according to the assumed  CSI error model. Firstly, the worst-case performance optimization approach was adopted in \cite{yu2019robust, Zhou2020robust} to design the active and passive beamforming jointly, by assuming the bounded CSI error model for partial channel uncertainty. The second approach adopted a probabilistic CSI error model (such as Gaussian distributed) and considered the average QoS performance \cite{you2019progressive, guo2019weighted, Zhao2020intelligent}. For example, in \cite{you2019progressive}, the time-varying IRS reflection based channel estimation was proposed and the resultant CSI errors over IRS reflecting elements were shown to follow the correlated Gaussian distribution due to the training with discrete IRS phase shifts in practice. The achievable rates of users in an IRS-aided multiuser system under the above CSI error model were characterized in \cite{Zhao2020intelligent}, where the reflection amplitude control of IRS (with/without the conventional phase-shift control) was further exploited to improve the system performance. Thirdly, the beamforming can be designed subject to a given QoS outage probability constraint for each user, i.e., the QoS performance needs to be above a certain threshold with a prescribed minimum probability (or its opposite maximum outage probability). In general, it is challenging to handle the outage probabilities since their closed-form expressions are usually difficult to obtain, especially in the multiuser context due to the inter-user interference. In \cite{zhou2020framework}, a Bernstein-type approach was adopted to solve the outage-constrained power minimization problem for an IRS-aided multiuser system, by assuming that the CSI errors are independent Gaussian random variables, which, however, cannot be applied to the case with other channel estimation error distributions. Therefore, it still remains unknown how to jointly optimize the active and passive beamforming for IRS-aided multiuser communications subject to QoS outage based constraints under correlated CSI errors and with discrete IRS phase shifts, which thus motivates this work.

In this paper, we consider the robust active and passive beamforming co-design in an IRS-aided multiple-input single-output (MISO) communication system, under the general case of correlated CSI errors. The active transmit precoding vectors at the AP and passive discrete phase shifts at the IRS are jointly optimized to minimize the AP's transmit power, subject to the outage probability constraints of the users and discrete uni-modular constraints on the reflection coefficients. By leveraging the results in \cite{Zhao2020intelligent} on the distribution of the IRS channel estimation errors, two new robust beamforming optimization algorithms are proposed for the single-user and multiuser cases, respectively. In particular, for the single-user case, we show that with given active transmit precoding vector and IRS phase shifts, the outage probability can be expressed in terms of the cumulative distribution function (cdf) of a non-central chi-square distribution with two degrees of freedom \cite{Abramowitz1972book}. Moreover, the outage probability is determined by the mean signal power (MSP) and variance of the received signal at the user. Since there is a nontrivial trade-off in tuning these two parameters to minimize the outage probability, we propose to maximize their weighted sum with the optimal weight found by one-dimensional search and present a weighted sum maximization (WSMax) algorithm. We show that when the weighting factor takes different values, the proposed WSMax algorithm reduces to three baseline algorithms, which correspond to maximizing the MSP, MSP to variance ratio (MVR), and MSP plus variance (MPV), respectively. For the multiuser case, since it is difficult to obtain closed-form expressions of the outage probabilities due to the inter-user interference, a novel two-stage constrained stochastic successive convex approximation (CSSCA) algorithm is proposed to solve the formulated multiuser robust beamforming optimization problem, where convex surrogate functions are iteratively constructed to replace the non-convex outage probability constraints and a two-stage procedure is devised to handle the discrete phase-shift constraints. Specifically, in the first stage, the IRS phase shifts are relaxed to continuous values, and the active and passive beamforming vectors are jointly optimized. Then, in the second stage, by quantizing the continuous phase shifts to discrete values and keeping them fixed, the active precoding vectors  are optimized to compensate for the outage performance loss caused by phase-shift quantization. Simulation results are presented to demonstrate the effectiveness of the proposed algorithms as compared to various benchmark schemes.

The rest of this paper is organized as follows. Section \ref{Section_system_model} introduces the system model and problem formulation. In Sections \ref{Section_single_user} and \ref{Section_multiuser}, we present the WSMax algorithm and two-stage CSSCA algorithm to solve the formulated problems in the single-user and multiuser cases, respectively. Section \ref{section_simulation} presents numerical results to evaluate the performance of the proposed algorithms and finally Section \ref{Section_conclusion} concludes the paper.

\emph{Notations}: Scalars, vectors and matrices are respectively denoted by lower/upper case, boldface lower-case and boldface upper-case letters. For an arbitrary matrix $\mathbf{A}$, $\mathbf{A}^T$, $\mathbf{A}^*$, $\mathbf{A}^{H}$ and $\mathbf{A}^{\dagger} $ denote its transpose, conjugate, conjugate transpose and pseudo-inverse, respectively. $\mathbf{A}^{-1}$ denotes the inverse of a square matrix $\mathbf{A}$ if it is invertible. $\mathbb{C}^{n\times m}$ denotes the space of $n\times m$ complex matrices and $\mathbb{R}^{n++}$ represents the space of $n\times1$ vectors with strictly positive real elements. For matrices $\mathbf{A} \in \mathbb{C}^{N_1 \times M}$ and $\mathbf{B} \in \mathbb{C}^{N_2 \times M}$, $[\mathbf{A};\mathbf{B}] \in \mathbb{C}^{(N_1+N_2)\times M}$ denotes row-wise concatenation of $\mathbf{A}$ and $\mathbf{B}$. $\|\cdot\|$ and $|\cdot|$ denote the Euclidean norm of a complex vector and absolute value of a complex number, respectively. 
$\mathcal{CN}(\mathbf{x},\bm{\Sigma})$ denotes the distribution of a circularly symmetric complex Gaussian (CSCG) random vector with mean vector $\mathbf{x}$ and covariance matrix $\bm{\Sigma}$; and $\sim$ stands for ``distributed as''.  $\mathcal{P}[\chi^2|_2, \lambda]$ denotes the cdf of a non-central chi-square distribution with two degrees of freedom and non-centrality parameter $\lambda$ \cite{Abramowitz1972book}. $Q_M(a,b)$ denotes the Marcum $Q$-function of real order $M > 0$ \cite{Abramowitz1972book}. For given numbers $\{x_1,\cdots,x_N\}$, $\textrm{diag}(x_1,\cdots,x_N)$ denotes a diagonal matrix with $\{x_1,\cdots,x_N\}$ being its diagonal elements, while $\textrm{diag}(\mathbf{A})$ denotes a vector that contains the diagonal elements of matrix $\mathbf{A}$. $\mathcal{F}^N$ is defined as the Cartesian product of $N$ identical sets each given by $\mathcal{F}$. The symbol $\jmath$ is used to represent $\sqrt{-1}$. For a complex number $x$, $\Re \{x\}$ ($\Im\{ x\}$) denotes its real (imaginary) part and $\angle x$ denotes its angle. $\mathbf{I}$ and $\mathbf{0}$ denote an identity matrix and an all-zero vector with appropriate dimensions, respectively. $\mathbb{E}\{\cdot\}$ denotes the statistical expectation. For given two sets $\mathcal{A}$ and $\mathcal{B}$, $\mathcal{A}\backslash \mathcal{B} \triangleq \{x| x\in\mathcal{A},x\notin \mathcal{B}\}$. The probability of an event $A$ is written as $\textrm{Pr}(A)$ and $\iint_{\mathbf{D}}(\cdot)$ denotes the double integral of a probability distribution function of a complex random variable over a disc $\mathbf{D}$ with certain center and radius in the two-dimensional plane.

\section{System Model and Problem Formulation} \label{Section_system_model}
\subsection{System Model}
\begin{figure}[!hhh] 
	\setlength{\abovecaptionskip}{-0.1cm} 
	\setlength{\belowcaptionskip}{-0.1cm} 
	\centering
	\scalebox{0.42}{\includegraphics{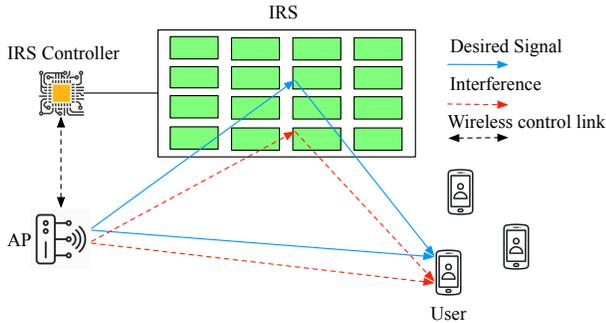}}
	\caption{An IRS-aided multiuser MISO downlink system.} 
	\label{fig:system_model}
\end{figure}

We consider an IRS-aided wireless system where an IRS equipped with $N$ reflecting elements is deployed to assist the communication between the AP (equipped with $M$ antennas) and $K$ single-antenna users (denoted by $\mathcal{K} \triangleq \{ 1,\cdots,K\}$), as shown in Fig. \ref{fig:system_model}. The IRS is attached with a smart controller that is able to adjust the reflection amplitude and/or phase shift of each reflecting element in real time and also communicates with the AP via a separate wireless link for coordinating transmission and exchanging information, such as CSI and IRS reflection coefficients \cite{Wu2019Magazine}. Denote by $\mathbf{h}_{d,k}^H \in \mathbb{C}^{1 \times M}$, $\mathbf{h}_{r,k}^H \in \mathbb{C}^{1 \times N}$ and $\mathbf{G} \in \mathbb{C}^{N \times M}$ the baseband equivalent channels from the AP to user $k$,  the IRS to user $k$, and the AP to the IRS, respectively. Let $\bm{\Theta} = \textrm{diag}(\phi_1,\cdots, \phi_N)$ denote the reflection-coefficient matrix at the IRS, where $\phi_n = a_n e^{\jmath \theta_n}$ ($n \in \mathcal{N} \triangleq\{1,\cdots,N \}$), $a_n\in[0,1]$ and $\theta_n \in [0,2\pi)$ are the reflection amplitude and phase shift of the $n$-th element, respectively. In this paper, the reflection amplitude of each element is set to $a_n = 1,\;\forall n \in\mathcal{N}$ to maximize the signal reflection for the ease of passive beamforming design and practical implementation. Moreover, we consider the practical constraint that the phase shift at each reflecting element only takes a finite number of discrete values \cite{Wu2019Discrete}. Let $Q$ denote the number of control bits for phase-shifting per IRS element. By assuming that the discrete phase-shift values are obtained by uniformly quantizing the interval $[0, 2\pi)$, we have $\phi_n \in \mathcal{F}_d \triangleq \{\phi_n| \phi_n = e^{\jmath \theta_n}, \theta_n \in \{0,\frac{2\pi}{Z},\cdots, \frac{2\pi(Z-1)}{Z}\}\}$, where $Z=2^Q$.

Then, the received signal at user $k$ can be expressed as
\begin{equation}
\begin{array}{l}
y_k =   (\mathbf{h}_{r,k}^H \mathbf{\Theta} \mathbf{G} + \mathbf{h}_{d,k}^H)\sum\limits_{j\in\mathcal{K}}\mathbf{w}_js_j + n_k,
\end{array}
\end{equation}
where $s_k$ represents the information symbol for user $k$ which is modeled as independent and identically distributed (i.i.d.) CSCG random variables with zero mean and unit variance; $\mathbf{w}_k \in \mathbb{C}^{M \times 1}$ denotes the transmit precoding vector for user $k$; $n_k$ denotes the i.i.d. complex additive white Gaussian noise (AWGN) at the receiver of user $k$ with zero mean and variance $\sigma_k^2$.
Thus, the signal-to-interference-plus-noise ratio (SINR) of user $k$ is given by
\begin{equation} \label{sinr_expression}
\textrm{SINR}_k = \frac{|(\mathbf{h}_{r,k}^H \mathbf{\Theta} \mathbf{G} + \mathbf{h}_{d,k}^H) \mathbf{w}_k|^2}{\sum\limits_{j\in\mathcal{K}\backslash k}|(\mathbf{h}_{r,k}^H \mathbf{\Theta} \mathbf{G} + \mathbf{h}_{d,k}^H)\mathbf{w}_j|^2+\sigma_k^2}.
\end{equation}

In practice, due to the limited channel training resources (e.g., power and time), perfect CSI is unlikely to obtain and CSI errors are inevitable. In this paper, we adopt the  time-varying reflection pattern based channel estimation method in \cite{Zhao2020intelligent, ZhaoGLOBECOM2020} to characterize the CSI error distribution. Specifically, let $\tilde{{\mathbf{H}}}_k \triangleq [\mathbf{h}_{d,k}^H; \mathbf{H}_k ] \in \mathbb{C}^{(N+1)\times M}$ denote the composite channel from the AP to user $k$, where $\mathbf{H}_{k} \triangleq \textrm{diag}(\mathbf{h}_{r,k}^H) \mathbf{G}$ denotes the cascaded AP-IRS-user $k$ channel, and define $\bar{\mathbf{H}}_k = [ \hat{\mathbf{h}}_{d,k}^H; \hat{\mathbf{H}}_k] $ with $ \hat{\mathbf{h}}_{d,k}$ and $\hat{\mathbf{H}}_k$ denoting the estimated channels. Then, by exploiting the uplink-downlink channel reciprocity and applying the least-square (LS) estimation, we have
\begin{equation}
\begin{aligned}
\bar{\mathbf{H}}_k & = \left(\frac{1}{\sqrt{p_{u,k}} s_{u,k}}\mathbf{Y}_{k} \mathbf{V}^{\dagger}\right)^H \\
& =\tilde{\mathbf{H}}_k +   \frac{1}{\sqrt{p_{u,k}} s_{u,k}}  (\mathbf{V}^{\dagger})^H \mathbf{N}_{u,k}^H,
\end{aligned}
\end{equation}
where $s_{u,k}$ denotes the uplink training symbol which is assumed to be $1$ without loss of optimality; $p_{u,k}$ denotes the uplink training power of user $k$; $\mathbf{N}_{u,k} = [\mathbf{n}^1_{u,k},\cdots,\mathbf{n}^{N_r}_{u,k}]$ with $\mathbf{n}^n_{u,k} \sim \mathcal{CN}(0,\varepsilon_{k}^2 \mathbf{I})$ denoting the uplink AWGN vector, $\varepsilon_{k}^2$ is the noise variance of user $k$ during channel training and $N_r$ denotes the total number of training symbols; $\mathbf{Y}_{k} \in \mathbb{C}^{M \times N_r}$ denotes the received uplink signal at the AP; $\mathbf{V} = [\tilde{\mathbf{v}}_1,\cdots,\tilde{\mathbf{v}}_{N_r}]$ and $ \tilde{\mathbf{v}}_n \triangleq [ 1, \mathbf{v}_n^T]^T$ ($\mathbf{v}_n= \textrm{diag}\{ \bm{\Theta}_n^*\}$) denotes the reflection pattern employed in the $n$-th training symbol duration. According to the analysis in \cite{Zhao2020intelligent}, the CSI error matrix $\Delta \tilde {\mathbf{H}}_k = \bar{\mathbf{H}}_k - \tilde{\mathbf{H}}_k$ satisfies $\mathbb{E} \{ \Delta \tilde {\mathbf{H}}_k\} = \mathbf{0}$ and $\mathbb{E} \{ \Delta \tilde {\mathbf{H}}_k \Delta \tilde {\mathbf{H}}_k^H  \}  = \frac{M\varepsilon_{k}^2}{p_{u,k}} (\mathbf{V} \mathbf{V}^H)^{\dagger}$. As can be seen, the CSI error matrix $\Delta \tilde {\mathbf{H}}_k$ is Gaussian distributed due to the fact that the channel noise is assumed to be CSCG and the adopted LS channel estimator only involves linear operations to the received uplink signals $\{\mathbf{Y}_k\}$. Besides, the elements in $\Delta \tilde {\mathbf{H}}_k$ are correlated if the training reflection vectors are non-orthogonal, i.e., $ (\mathbf{V}^{\dagger})^H \mathbf{V}^{\dagger} \neq \mathbf{I}$. This usually occurs in practice when discrete phase shifts are used and/or $N_r > N + 1$.\footnote{Note that there are only $N+1$ elements in the orthogonal basis of the $(N+1)$-dimensional space. Therefore, if the training duration $N_r$ is larger than $N+1$, it is inevitable that at least one training reflection vector would be non-orthogonal to the others.} Similar to \cite{Zhao2020intelligent}, we assume in this paper that 1) $\mathbf{V}$ is given (e.g., $\{\tilde{\mathbf{v}}_n \}$ can be chosen to be the columns of the quantized DFT matrix or truncated Hadamard matrix according to the value of $Q$ \cite{you2019progressive}), 2) the uplink training powers $\{p_{u,k}\}$ are known at the AP, and 3) the uplink channel noise variances $\{\varepsilon_{k}^2\}$ can be effectively estimated by off-the-shelf methods, see e.g., \cite{Das2012} and the references therein, thus the CSI error covariance matrices $\mathbb{E} \{ \Delta \tilde {\mathbf{H}}_k \Delta \tilde {\mathbf{H}}_k^H  \}, \forall k \in \mathcal{K}$ are fixed and known at the AP. Note that although calculating $\mathbb{E} \{ \Delta \tilde {\mathbf{H}}_k \Delta \tilde {\mathbf{H}}_k^H  \}$ involves the matrix inversion operation, the required computational complexity can be practically ignored since $\mathbf{V}$ is known in advance and this calculation only needs to be performed once without the need of knowing any channel statistics.

\begin{figure}[!hhh] 
	\setlength{\abovecaptionskip}{-0.1cm} 
	\setlength{\belowcaptionskip}{-0.1cm} 
	\centering
	\scalebox{0.45}{\includegraphics{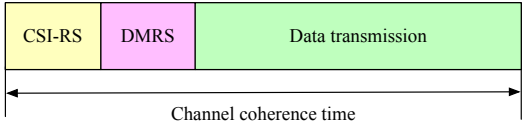}}
	\caption{Transmission protocol in the considered IRS-aided communication system.} 
	\label{fig:transmission_protocol}
\end{figure}

In this paper, we assume that the AP can only obtain the full CSI $\bar{\mathbf{H}}$ imperfectly, while the users can obtain their individual perfect AP-user effective CSI, i.e., $\{(\mathbf{h}_{r,k}^H \mathbf{\Theta} \mathbf{G} + \mathbf{h}_{d,k}^H) \mathbf{w}_k\}$, which includes the transmit precoding vectors at the AP and the reflection phase shifts at the IRS. As a result, the users can accurately evaluate the SINR in \eqref{sinr_expression}. To justify this assumption, we illustrate in Fig. \ref{fig:transmission_protocol} a practical transmission protocol, which is a simplified version of the 5G NR protocol specified in \cite{3GPPRelease15}. More specifically, there are two types of training symbols, also known as reference signals (RSs), in the proposed transmission protocol, i.e., the channel state information-RS (CSI-RS) and the demodulation RS (DMRS), where the former is used to estimate the full CSI for joint active and passive beamforming design, while the latter is employed to estimate the effective CSI for demodulation. Since the dimension of the effective CSI (a complex scalar for each user) is much smaller than that of the full CSI and the DMRS is beamformed (with the designed $\{\mathbf{w}_k\}$ and $\bm{\Theta}$), it is reasonable to assume that the estimated effective CSI is much more accurate than the estimated full CSI. Therefore, in this paper, we focus on mitigating the outage caused by the full CSI error at the AP side, and assume perfect effective CSI at the user side (with the optimized active and passive beamformers). On the other hand, even if the estimated effective CSI is imperfect, we can always impose more conservative constraints on the users' outage probabilities, such that the resulting negative effects can be effectively compensated. Note that this assumption is in consistent with the existing literature on outage-constrained robust beamforming design, see e.g.,  \cite{Wang2014Outage, zhou2020framework}.\footnote{Note that in  \cite{Zhao2020intelligent, Ozdogan2019TCOM, Jung2020TWC}, there is an additional interference term in the SINR expression, which is caused by imperfect effective CSI estimated at the users. In our considered system, according to the transmission protocol in Fig. \ref{fig:transmission_protocol}, the estimated effective CSI is much more accurate than the estimated full CSI, thus it will have little impact on the outage probabilities. As a result, we ignore the impacts of the effective CSI errors in this paper and the resulting interference is also not considered.}

\begin{figure}[!hhh] 
	\setlength{\abovecaptionskip}{-0.1cm} 
	\setlength{\belowcaptionskip}{-0.1cm} 
	\centering
	\scalebox{0.41}{\includegraphics{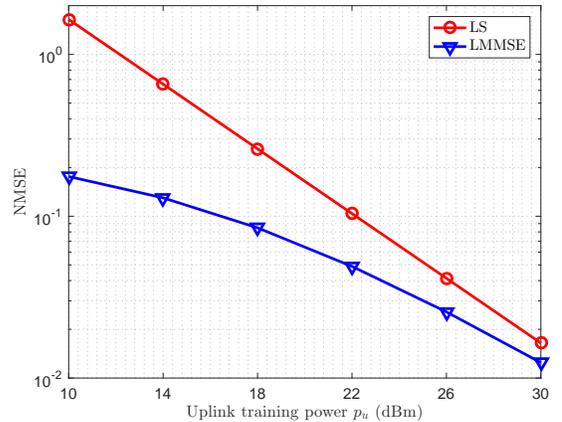}}
	\caption{Performance comparison between the LS and LMMSE estimators.} \label{channel_estimation_comparison}
\end{figure}

\newtheorem{remark}{Remark}
	\begin{remark}
	\emph{Generally, the linear minimum mean square error (LMMSE) estimation method can achieve better estimation accuracy than the LS method \cite{Oppenheim2015}. In our case, the LMMSE estimation of $\tilde{\mathbf{H}}_k$ can be expressed as \cite{Kay1993}
		\begin{equation} \label{LMMSE_estimator}
			\begin{aligned}
				\bar{\mathbf{H}}_k^{\textrm{LMMSE}}  = & \sqrt{p_{u,k}} \mathbf{C}_{H_k}^H \mathbf{V} (p_{u,k} \mathbf{V}^H \mathbf{C}_{H_k}\mathbf{V} + M \varepsilon_{k}^2 \mathbf{I})^{-1} \\
				& \times (\mathbf{Y}_k - \mathbb{E}\{\mathbf{Y}_k \})^H+ \mathbb{E}\{\tilde{\mathbf{H}}_k \}\\
				 =  & \sqrt{p_{u,k}} \mathbf{C}_{H_k}^H \mathbf{V} (p_{u,k} \mathbf{V}^H \mathbf{C}_{H_k}\mathbf{V} + M \varepsilon_{k}^2 \mathbf{I})^{-1}  \\
				 & \times (\mathbf{Y}_k - \sqrt{p_{u,k}}\mathbb{E}\{\tilde{\mathbf{H}}_k^H\} \mathbf{V}  )^H+ \mathbb{E}\{\tilde{\mathbf{H}}_k \},
			\end{aligned}
		\end{equation}
		where $\mathbf{C}_{H_k} \triangleq \mathbb{E}\{(\tilde{\mathbf{H}}_k- \mathbb{E}\{\tilde{\mathbf{H}}_k \})(\tilde{\mathbf{H}}_k- \mathbb{E}\{\tilde{\mathbf{H}}_k \})^H \}$. Accordingly, the corresponding CSI error covariance matrix is given by
		\begin{equation} \label{error_covariance}
			\begin{aligned}
			 \mathbb{E}  \{ \Delta \tilde  {\mathbf{H}}_k \Delta \tilde {\mathbf{H}}_k^H  \}  &
			= \mathbf{C}_H  - p_{u,k}\mathbf{C}_H \mathbf{V}  \\
			& \times (p_{u,k} \mathbf{V}^H \mathbf{C}_{H_k}\mathbf{V} + M \varepsilon_{k}^2 \mathbf{I})^{-1}  \mathbf{V}^H \mathbf{C}_H.
			\end{aligned}
		\end{equation}
	Fig. \ref{channel_estimation_comparison} shows the performance comparison between the LS and LMMSE estimators, where the simulation parameters specified in Section \ref{section_simulation} are adopted and the normalized mean square error (NMSE) is considered as the performance metric, which is defined as $\textrm{NMSE} \triangleq \sum\nolimits_{k \in \mathcal{K}}\|\bar{\mathbf{H}}_k - \tilde{\mathbf{H}}_k\|^2 / \sum\nolimits_{k \in \mathcal{K}} \|\tilde{\mathbf{H}}_k\|^2$. It is observed that the LMMSE estimator outperforms the LS estimator, especially in the low-$p_u$ regime.
		 However, as the price paid for better performance, the LMMSE estimator requires complex matrix inversion operations and some statistical knowledge about the channel, such as the channel mean values $\mathbb{E}\{\tilde{\mathbf{H}}_k\},\forall k \in \mathcal{K}$ and covariance matrices $\mathbf{C}_{H_k},\forall k \in \mathcal{K}$. These additional costs may become overwhelming for a practical IRS-aided communication system, especially when $N$ becomes large.\footnote{In Fig. \ref{channel_estimation_comparison}, $\mathbb{E}\{\tilde{\mathbf{H}}_k\}$ and $\mathbf{C}_{H_k},\forall k \in \mathcal{K}$ are obtained via extensive sample averaging.} Therefore, for ease of practical implementation, we adopt the LS estimation method in this paper, which is simpler and more computational friendly. Nevertheless, we note that the proposed algorithms (as will be introduced later) are still applicable when other channel estimation methods are employed as long as the statistical information of the  CSI error matrix can be obtained. For example, if we use the LMMSE channel estimation method and  assume that $\mathbb{E}\{\tilde{\mathbf{H}}_k\}$ and $\mathbf{C}_{H_k},\forall k \in \mathcal{K}$ are known, then the CSI error matrix is still complex Gaussian distributed (since in LMMSE, the estimated channel is a linear transformation of the received signal corrupted by CSCG noise) and the AP can also calculate the CSI error covariance matrix according to \eqref{error_covariance}. Therefore, the proposed algorithms are still applicable in this case and  the difference is that the CSI error covariance matrix under LMMSE is different. However, since $\mathbb{E}\{\tilde{\mathbf{H}}_k\}$ and $\mathbf{C}_{H_k},\forall k \in \mathcal{K}$ are difficult to obtain in practice and they may also change over time, further investigation of the channel statistics estimation, the corresponding robust beamforming design and performance comparison is required, which is not considered in this paper.}
	\end{remark}

\subsection{Problem Formulation} \label{CE_model}
In this paper, we aim to minimize the AP transmit power subject to the individual SINR outage probability constraints at the users as well as the discrete reflection coefficient constraints, by jointly optimizing the active transmit precoders at the AP and passive phase shifts at the IRS. Accordingly, the optimization problem is formulated as
\begin{subequations} \label{outage_problem}
\begin{align}
 \min\limits_{\{\mathbf{w}_k\},\;\bm{\Theta}} \; & \sum\limits_{k \in\mathcal{K}} \| \mathbf{w}_k\|^2\\
 \textrm{s.t.}\; & \textrm{Pr}( \textrm{SINR}_k < \eta_k ) \leq \epsilon_k,\;\forall k \in \mathcal{K}, \label{outage_constraint}\\
& \phi_n \in \mathcal{F}_d,\;\forall n \in \mathcal{N}.
\end{align}
\end{subequations}
The outage probability constraints in \eqref{outage_constraint} guarantee the users' QoS, i.e., the probability of each user that can successfully decode its message at a transmission rate of $\log_2(1+\eta_k)$ is no less than $1-\epsilon_k$. This type of design is practically useful for, e.g., delay-sensitive or low-latency applications, where the system is required to provide a prescribed communication rate with high probability.

Different from the prior works \cite{Zhao2020intelligent, Wu2019Discrete, ZhaoGLOBECOM2020}, solving problem \eqref{outage_problem} requires efficiently handling the outage probabilities given in \eqref{outage_constraint}, which are difficult to be characterized, especially in the multiuser context due to the inter-user interference. Besides, since the CSI errors are correlated in general, the discrete IRS phase shifts in $\mathbf{\Theta}$ and active precoding vectors $\{\mathbf{w}_k \}$ are intricately coupled in the constraints given in \eqref{outage_constraint} and thus difficult to be jointly optimized. In the next two sections, we present useful techniques to deal with the above difficulties and propose efficient algorithms to solve problem \eqref{outage_problem} sub-optimally in the single-user and multiuser cases, respectively.

\section{Single-User System} \label{Section_single_user}
In this section, we consider the single-user case, i.e., $K = 1$, to draw useful insights into the effect of IRS phase shifts on the outage probability. In this case, multiuser interference does not exist, therefore we can simply drop the user subscript $k$ and ignore the multiuser interference term in \eqref{sinr_expression}, which leads to the following optimization problem:
\begin{equation} \label{single_user_problem}
	\begin{aligned}
	\min\limits_{\mathbf{w},\; \bm{\Theta}} \; &  \| \mathbf{w}\|^2\\
	\textrm{s.t.}\; & \textrm{Pr} \left( {|(\mathbf{h}_{r}^H \mathbf{\Theta} \mathbf{G} + \mathbf{h}_{d}^H) \mathbf{w}|^2} < \sigma^2 \eta \right) \leq \epsilon, \\
&	 \phi_n \in \mathcal{F}_d,\;\forall n \in \mathcal{N}.
	\end{aligned}
\end{equation}
Let $\mathbf{v}= \textrm{diag}(\bm{\Theta}^*)$, problem \eqref{single_user_problem} is equivalent to 
\begin{equation} \label{single_user_problem_eq}
	\begin{aligned}
	\min\limits_{\mathbf{w},\;\mathbf{\mathbf{v}}} \; &  \| \mathbf{w}\|^2\\
	\textrm{s.t.}\; & \textrm{Pr}\left({|( \mathbf{v}^H \mathbf{H} + \mathbf{h}_{d}^H) \mathbf{w}|^2} <  \sigma^2\eta \right)  \leq \epsilon, \\
	&   v_n \in \mathcal{F}_d,\;\forall n \in \mathcal{N}.
	\end{aligned}
\end{equation}
We observe that the outage probability is a decreasing function of the downlink transmit power $\|\mathbf{w} \|^2$, i.e., for any $\mathbf{w}_1$ that satisfies $\|\mathbf{w}_1 \|^2=p_1 $, we can always let $\mathbf{w}_2 = \sqrt{\frac{p_2}{p_1}} \mathbf{w}_1$ ($p_2 \geq p_1$) such that $\textrm{Pr}\left({|( \mathbf{v}^H \mathbf{H} + \mathbf{h}_{d}^H) \mathbf{w}_1|^2} <  \sigma^2\eta \right) \geq \textrm{Pr}\left({|( \mathbf{v}^H \mathbf{H} + \mathbf{h}_{d}^H) \mathbf{w}_2|^2} <  \sigma^2\eta \right)$ holds. Therefore, instead of minimizing the AP transmit power $\|\mathbf{w}\|^2$ under the outage probability constraint as in \eqref{single_user_problem_eq}, we can alternatively minimize the outage probability with given $\|\mathbf{w}\|^2$ and then search for the appropriate $\|\mathbf{w}\|^2$ with which the achieved outage probability is less than or equal to the outage probability target $\epsilon$. As a result, problem \eqref{single_user_problem_eq} can be tackled by solving a sequence of outage probability minimization problems, i.e.,
\begin{equation} \label{outage_minimize}
	\begin{aligned}
	\min \limits_{\mathbf{w},\; \mathbf{\mathbf{v}}} \; &  \textrm{Pr}\left({|( \mathbf{v}^H \mathbf{H} + \mathbf{h}_{d}^H) \mathbf{w}|^2} < \sigma^2\eta \right)\\
	\textrm{s.t.}\; & \| \mathbf{w}\|^2 \leq p, \\
	 & v_n \in \mathcal{F}_d,\;\forall n \in \mathcal{N},
	\end{aligned}
\end{equation}
where the minimum required $p$ can be found via bisection search. Note that after solving one instance of problem \eqref{outage_minimize} with given $p$, we should decrease or increase $p$ according to the achieved outage probability, thus problem \eqref{outage_minimize} is not explicitly related to $\epsilon$.

In the following, we focus on problem \eqref{outage_minimize} with given $p$ and propose an efficient algorithm, called the WSMax algorithm, to solve it. In particular, we first derive a closed-form expression of the outage probability based on the statistics of the CSI errors and show that minimizing the outage probability is equivalent to optimizing the MSP and variance of the received signal at the user. Since there is a trade-off in tuning these two parameters to minimize the outage probability, we propose to solve problem \eqref{outage_minimize} by maximizing the weighted sum of the MSP and variance via the penalty dual decomposition (PDD) method \cite{ShiPDD2020} and finding the optimal weighting factor by one-dimensional search. Furthermore, we demonstrate that the proposed WSMax algorithm contains three baseline algorithms (namely, the MVR, MPV and MSP maximization algorithms)  as special cases when the weighting factor takes different values.

\subsection{Problem Transformation}
First, by letting $\tilde{\mathbf{v}} = [1,\mathbf{v}^T]^T$, we have the following equivalent form of problem \eqref{outage_minimize}:
\begin{equation} \label{w_subproblem}
\begin{aligned}
\min \limits_{\mathbf{w},\; \mathbf{v}} \; &  \textrm{Pr}\left( {| \tilde{\mathbf{v}}^H \tilde{\mathbf{H}} \mathbf{w}|^2} < \sigma^2\eta \right) \\
\textrm{s.t.}\; & \| \mathbf{w}\|^2 \leq p, \\
 & v_n \in \mathcal{F}_d,\;\forall n \in \mathcal{N}.
\end{aligned}
\end{equation}
It can be seen that since $\tilde{\mathbf{H}}= \bar{\mathbf{H}}  - \Delta \tilde {\mathbf{H}} $ and $\Delta \tilde {\mathbf{H}}$ is complex Gaussian distributed with $\mathbb{E} \{ \Delta \tilde {\mathbf{H}}\} = \mathbf{0}$ and $\mathbb{E} \{ \Delta \tilde {\mathbf{H}} \Delta \tilde {\mathbf{H}}^H  \}  = \frac{M\varepsilon^2}{p_{u}} (\mathbf{V} \mathbf{V}^H)^{\dagger}$, $z= \tilde{\mathbf{v}}^H \tilde{\mathbf{H}} \mathbf{w}$ is a complex random variable with mean $\tilde{\mathbf{v}}^H\bar{\mathbf{H}} \mathbf{w}$ and variance $\mathbb{E}\{ \mathbf{w}^H \Delta \tilde{\mathbf{H}}^H \tilde{\mathbf{v}} \tilde{\mathbf{v}}^H \Delta \tilde{\mathbf{H}} \mathbf{w} \}$. Moreover, since $\Delta \tilde{\mathbf{H}} =\frac{1}{\sqrt{p_u}} (\mathbf{V}^{\dagger})^H \mathbf{N}_u^H$, the variance $\mathbb{E}\{ \mathbf{w}^H \Delta \tilde{\mathbf{H}}^H \tilde{\mathbf{v}} \tilde{\mathbf{v}}^H \Delta \tilde{\mathbf{H}} \mathbf{w} \}$ can be expressed as
	\begin{equation}
\begin{aligned}
&	\mathbb{E}\{ \mathbf{w}^H \Delta \tilde{\mathbf{H}}^H  \tilde{\mathbf{v}} \tilde{\mathbf{v}}^H \Delta \tilde{\mathbf{H}} \mathbf{w} \} \\
	& =  \frac{1}{p_u} \mathbb{E}\{ \mathbf{w}^H   \mathbf{N}_u \mathbf{V}^{\dagger}  \tilde{\mathbf{v}} \tilde{\mathbf{v}}^H  (\mathbf{V}^{\dagger})^H \mathbf{N}_u^H \mathbf{w} \} \\
	& =  \frac{\varepsilon^2}{p_u}  \tilde{\mathbf{v}}^H  (\mathbf{V}^{\dagger})^H \mathbf{V}^{\dagger}  \tilde{\mathbf{v}}  \mathbf{w}^H \mathbf{w} \\
&	= p  \tilde{\mathbf{v}}^H  \bar{\mathbf{V}} \tilde{\mathbf{v}},
\end{aligned}
\end{equation}
where $\bar{\mathbf{V}} = \frac{\varepsilon^2}{p_u}  (\mathbf{V}^{\dagger})^H \mathbf{V}^{\dagger} $. Therefore, problem \eqref{w_subproblem} can be equivalently rewritten as  
\begin{equation} \label{w_subproblem_eq}
\begin{aligned}
\min \limits_{\mathbf{w},\;\mathbf{v}} \; & \iint_{\mathbf{D}} \mathcal{CN}(z; \tilde{\mathbf{v}}^H \bar{\mathbf{H}} \mathbf{w}, p \tilde{\mathbf{v}}^H  \bar{\mathbf{V}}  \tilde{\mathbf{v}} )  \\
\textrm{s.t.}\; & \| \mathbf{w}\|^2 \leq p, \\
&  v_n \in \mathcal{F}_d,\;\forall n \in \mathcal{N},
\end{aligned}
\end{equation}
where $\mathbf{D}$ is a disc centered at zero with radius $\sqrt{\sigma^2 \eta}$ in the complex plane. Then, we can see that $\Re (z)$ and $\Im(z)$ are two independently and normally distributed random variables with mean values $\Re\{  \tilde{\mathbf{v}}^H \bar{\mathbf{H}} \mathbf{w}\}$ and $\Im \{ \tilde{\mathbf{v}}^H \bar{\mathbf{H}} \mathbf{w}\}$, respectively, and variance $\frac{1}{2} p \tilde{\mathbf{v}}^H  \bar{\mathbf{V}}  \tilde{\mathbf{v}}$. Therefore, the double integral in the objective function of problem \eqref{w_subproblem_eq} is given by
\begin{equation} \label{eval_outage}
\begin{aligned}
& \iint_{\mathbf{D}} \mathcal{CN}(z; \tilde{\mathbf{v}}^H \bar{\mathbf{H}} \mathbf{w},  p\tilde{\mathbf{v}}^H \bar{\mathbf{V}}  \tilde{\mathbf{v}} ) \\
& = \mathcal{P} \left( \frac{\eta\sigma^2}{\frac{1}{2} p \tilde{\mathbf{v}}^H \bar{\mathbf{V}} \tilde{\mathbf{v}} } \bigg|_2,\frac{|\tilde{\mathbf{v}}^H \bar{\mathbf{H}} \mathbf{w}|^2}{  \frac{1}{2}  p \tilde{\mathbf{v}}^H  \bar{\mathbf{V}} \tilde{\mathbf{v}}  } \right).
 \end{aligned}
\end{equation}
Since $\mathcal{P}[\chi^2|_2,\lambda] = 1-Q_1(\sqrt{\lambda}, \chi)$ and $Q_1(\sqrt{\lambda}, \chi)$ is a strictly increasing function of $\lambda$ for all $\chi >0 $ \cite{Sun2010TIT}, we have $\mathcal{P}[\chi^2|_2, \lambda_1 ] \leq \mathcal{P}[\chi^2|_2, \lambda_2]$ for any $\lambda_2 \geq \lambda_1$; thus, the optimal $\mathbf{w}$ of problem \eqref{w_subproblem_eq} is the maximum-ratio transmission (MRT) beamformer $\mathbf{w} = \sqrt{p}{(\tilde{\mathbf{v}}^H \bar{\mathbf{H}})^H}/{\|\tilde{\mathbf{v}}^H \bar{\mathbf{H}}\|}$ and the power constraint in problem \eqref{w_subproblem_eq} must be satisfied with equality. Consequently, problem \eqref{w_subproblem_eq} is further equivalent to
\begin{equation} \label{v_subproblem_eq2}
\begin{aligned}
\min \limits_{\mathbf{v}} \; & \mathcal{C} (\mathbf{v};p) \triangleq \mathcal{P} \left( \frac{\eta\sigma^2}{\frac{1}{2} p \tilde{\mathbf{v}}^H \bar{\mathbf{V}} \tilde{\mathbf{v}} } \bigg|_2,\frac{\tilde{\mathbf{v}}^H \bar{\mathbf{H}}\bar{\mathbf{H}}^H \tilde{\mathbf{v}}}{  \frac{1}{2} \tilde{\mathbf{v}}^H  \bar{\mathbf{V}} \tilde{\mathbf{v}}  } \right)\\
\textrm{s.t.}\; & v_n \in \mathcal{F}_d,\;\forall n \in \mathcal{N},
\end{aligned}
\end{equation}
where the transmit power constraint $\|\mathbf{w}\|^2 \leq p$ in problem \eqref{w_subproblem_eq} is safely ignored since $\|\mathbf{w}\|^2 = p$ is automatically satisfied by employing the MRT beamformer.
	
Note that the MSP $\tilde{\mathbf{v}}^H \bar{\mathbf{H}}\bar{\mathbf{H}}^H \tilde{\mathbf{v}}$ and variance $\tilde{\mathbf{v}}^H \bar{\mathbf{V}} \tilde{\mathbf{v}}$ are both related to $\tilde{\mathbf{v}}$ (the constant $\frac{1}{2}$ and $p$ can be ignored without loss of generality). Moreover, we can observe from \eqref{v_subproblem_eq2} that in order to minimize the outage probability $\mathcal{C} (\mathbf{v};p)$, we should maximize the non-centrality parameter (i.e., the MVR) $\frac{\tilde{\mathbf{v}}^H \bar{\mathbf{H}}\bar{\mathbf{H}}^H \tilde{\mathbf{v}}}{ \tilde{\mathbf{v}}^H  \bar{\mathbf{V}} \tilde{\mathbf{v}}  }$ and variance $\tilde{\mathbf{v}}^H \bar{\mathbf{V}} \tilde{\mathbf{v}}$ simultaneously, based on the fact that $\mathcal{P}[\chi^2|_2,\lambda]$ is a deceasing  function of $\lambda$ and an increasing function of $\chi^2$ for all $\lambda>0$ and $\chi>0$. However, the MVR and variance cannot be maximized at the same time in general via tuning $\tilde{\mathbf{v}}$ since larger variance usually leads to smaller MVR, which renders problem \eqref{v_subproblem_eq2} difficult to solve. Finally, it is worth noting that minimizing the variance $\tilde{\mathbf{v}}^H \bar{\mathbf{V}} \tilde{\mathbf{v}}$ does not lead to the minimum outage probability in general, which may be counter-intuitive at the first glance. This is because in our case, flexible variance is needed to trade-off between maximizing MSP and minimizing variance for outage probability minimization.

\subsection{Proposed WSMax Algorithm}
\begin{figure*}[!hhh] 
	\setlength{\abovecaptionskip}{-0.1cm} 
	\setlength{\belowcaptionskip}{-0.1cm} 
	\centering
	\scalebox{0.4}{\includegraphics{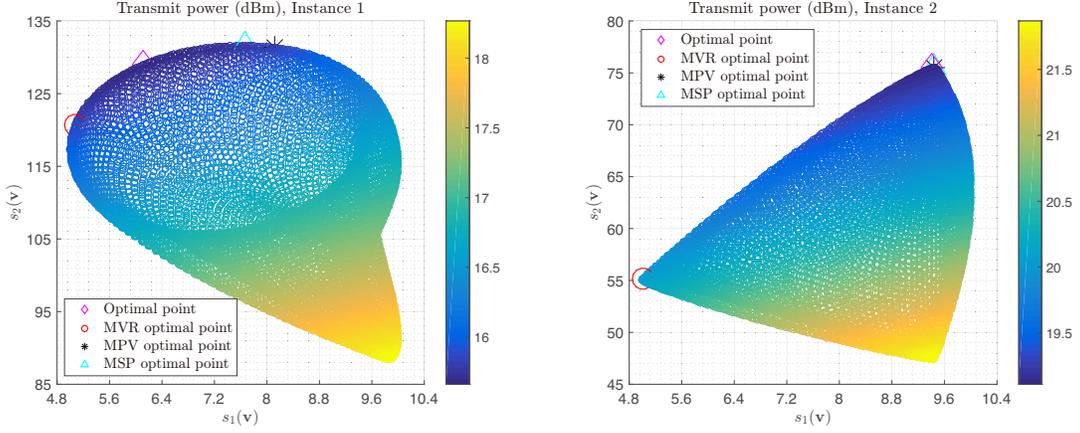}}
	\caption{Numerical example of the MSP-variance region for the continuous phase-shift case.} 
	\label{fig:region_continuous}
\end{figure*}

\begin{figure*}[!hhh] 
	\setlength{\abovecaptionskip}{-0.1cm} 
	\setlength{\belowcaptionskip}{-0.1cm} 
	\centering
	\scalebox{0.4}{\includegraphics{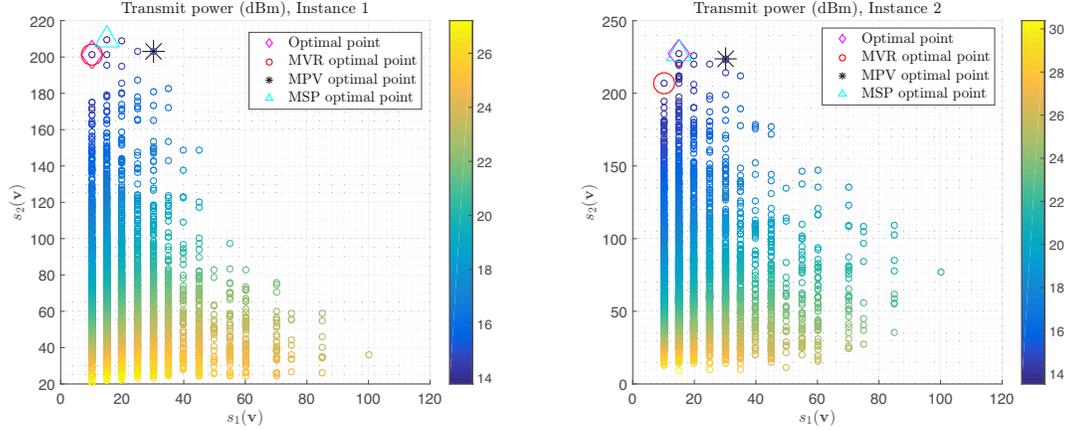}}
	\caption{Numerical example of the MSP-variance region for the discrete phase-shift case.} 
	\label{fig:region_discrete}
\end{figure*}

In this subsection, we propose the WSMax algorithm to efficiently solve problem \eqref{v_subproblem_eq2} sub-optimally. First, in order to gain insight into the optimal solution of problem \eqref{v_subproblem_eq2}, we define the following MSP-variance region.
\newtheorem{definition}{Definition} 
\begin{definition}[MSP-variance region]
\emph{
The MSP-variance region of the considered IRS-aided single-user MISO downlink system with correlated CSI errors is given by
\begin{equation} \label{region_definition}
\begin{aligned}
\mathcal{S} \triangleq \{(S_1,S_2): & S_1 =  \tilde{\mathbf{v}}^H \bar{\mathbf{V}} \tilde{\mathbf{v}} \triangleq s_1(\mathbf{v}), \\
& S_2 = {\tilde{\mathbf{v}}^H \bar{\mathbf{H}}\bar{\mathbf{H}}^H \tilde{\mathbf{v}}} \triangleq s_2(\mathbf{v}), \mathbf{v}\in \mathcal{F}_d^N \}.
\end{aligned}
\end{equation}
}
\end{definition}
\noindent We observe that $\mathcal{S}$ is a compact set since it is finite and $\mathcal{S}$ is in general disconnected due to the discrete phase shifts at the IRS. For the continuous phase-shift case, since the set of feasible IRS phase shifts $\mathcal{F}^N$ ($\mathcal{F} \triangleq \{\phi_n| |\phi_n|=1\}$) is compact and $s_i(\mathbf{v}), i\in  \{ 1,2 \}$ are continuous functions of $\mathbf{v}$ by definition, $\mathcal{S}$ is also a compact set since it is the image of a continuous mapping from $\mathcal{F}^N$ \cite[Theorem 4.14]{rudin1964principles}. Besides, $\mathcal{S}$ is not a normal set since it does not satisfy that for any point $\mathbf{s} \in \mathcal{S}$, all $\bar{\mathbf{s}} \in \mathbb{R}^{2++}$ with $\bar{\mathbf{s}} \leq \mathbf{s}$ also satisfy $\bar{\mathbf{s}} \in \mathcal{S}$ \cite{tuy2000monotonic}, which is reasonable since either $s_1(\mathbf{v}) = 0$ or $s_2(\mathbf{v}) = 0$ is unlikely to be true due to the uni-modular constraints on the IRS phase-shift vector $\mathbf{v}$. Therefore, $\mathcal{S}$ is generally non-convex and it is difficult to characterize the MSP-variance region as both the MSP and variance are nonlinearly coupled with  $\mathbf{v}$. In Figs. \ref{fig:region_continuous} and \ref{fig:region_discrete}, we illustrate some numerical examples of the MSP-variance region for both continuous and discrete phase-shift cases based on the simulation setup in Section \ref{section_simulation}, where 
higher color temperature means larger transmit power and vice versa. For the continuous phase-shift case, we set $N=2$ and $N_r = 4$, while for the discrete phase-shift case, we set $Q=1$, $N=12$ and $N_r = N+1$.  As can be observed, the MVR, MPV and MSP optimal points are not necessarily the optimal point (i.e., achieving the lowest transmit power with the given outage probability constraint) for both continuous and discrete phase-shift cases, and the MSP-variance region is non-convex in general.


To proceed, we further define the upper boundary of $\mathcal{S}$ as follows.
\begin{definition}[Upper boundary point]
	\emph{
A point $\mathbf{s} = (S_1, S_2) \in \mathbb{R}^{2++}$ is called an upper boundary point of $\mathcal{S}$ if $\mathbf{s} \in \mathcal{S}$ while the set $\{\bar{\mathbf{s}} = (\bar{S}_1,\bar{S}_2)\in \mathbb{R}^{2++}: \bar{S}_1 = S_1, \bar{S}_2> S_2 \} \subseteq \mathbb{R}^{2++}\backslash \mathcal{S}$.
	}
\end{definition}
\noindent 
Based on this definition, we can infer that the optimal solution of problem \eqref{v_subproblem_eq2} must lie on the upper boundary of $\mathcal{S}$ since with fixed $s_1(\mathbf{v})$, larger $s_2(\mathbf{v})$ always leads to smaller outage probability (or equivalently, lower transmit power), which can also be observed from Figs. \ref{fig:region_continuous} and \ref{fig:region_discrete}. Therefore, we propose to maximize the weighted sum of the MSP and variance (i.e., $s_2(\mathbf{v})$ and $s_1(\mathbf{v})$), which leads to the following optimization problem:
\begin{equation} \label{weighted_sum_problem}
\begin{aligned}
\max\limits_{\mathbf{v}} \;& {\tilde{\mathbf{v}}^H \bar{\mathbf{H}}\bar{\mathbf{H}}^H \tilde{\mathbf{v}}} + \omega \tilde{\mathbf{v}}^H \bar{\mathbf{V}} \tilde{\mathbf{v}} \\
\textrm{s.t.}\;  & v_n \in \mathcal{F}_d,\;\forall n \in \mathcal{N},
\end{aligned}
\end{equation}
where $\omega$ denotes the weighting factor which is a real number and can be positive, negative or zero. It is noteworthy that the AP transmit power $p$ is not involved in problem \eqref{weighted_sum_problem}, therefore problem \eqref{weighted_sum_problem} does not need to be solved under different values of $p$, i.e., the bisection search over $p$ can be conducted with the obtained $\mathbf{v}$ after solving \eqref{weighted_sum_problem}.

Problem \eqref{weighted_sum_problem} is generally a non-convex quadratic programming problem with discrete constraints and it can be efficiently solved by employing a similar PDD-based algorithm as that in \cite{zhao2019intelligent}. However, since $\omega$ can be negative and the objective function of problem \eqref{weighted_sum_problem} may contain both convex and concave components, certain modifications need to be made which will become clear later. In the following, we present the modified PDD-based algorithm to solve problem \eqref{weighted_sum_problem}, which mainly consists of two loops. In the inner loop, the block successive upper-bound minimization (BSUM) method is employed to iteratively optimize the primal variables in different blocks, while in the outer loop, we update the dual variable and penalty parameter. Specifically, similar to \cite{zhao2019intelligent}, we introduce an auxiliary variable $\mathbf{u}=[u_1,\cdots,u_N]^T$ which satisfies $\mathbf{u} = \mathbf{v}$ and an additional constraint $\|\mathbf{v}\|^2 \leq N$, then the augmented Lagrangian problem of \eqref{weighted_sum_problem} can be written as follows:
\begin{equation}
\begin{aligned}
\min\limits_{\mathbf{v},\;\mathbf{u}} \;& -\tilde{\mathbf{v}}^H \bar{\mathbf{H}}\bar{\mathbf{H}}^H \tilde{\mathbf{v}} - \omega \tilde{\mathbf{v}}^H \bar{\mathbf{V}} \tilde{\mathbf{v}} +\frac{1}{2\rho} \|\mathbf{v}-\mathbf{u}+\rho \bm{\lambda} \|^2\\
\textrm{s.t.}\;  &\|\mathbf{v}\|^2 \leq N,\\
& u_n \in \mathcal{F}_d,\;\forall n \in \mathcal{N},
\end{aligned}
\end{equation}
where $\rho$ is the penalty parameter and $\bm{\lambda} = [\lambda_1, \cdots,  \lambda_N]^T$ denotes the dual variable vector associated with the constraint $\mathbf{v} = \mathbf{u}$.
Then, we alternately optimize $\mathbf{v}$ and $\mathbf{u}$ in the inner loop. In the $\mathbf{v}$-optimization step, we have the following problem: 
\begin{equation} \label{v_subproblem}
\begin{aligned}
\min\limits_{\mathbf{v}} \;& \mathbf{v}^H\left(\frac{1}{2\rho}  \mathbf{I} - \mathbf{U} \bm{\Sigma}^-\mathbf{U}^H \right) \mathbf{v}  \\
&+ 2  \Re\left\{\mathbf{v}^H \left(\frac{1}{2}\bm{\lambda}-\frac{1}{2\rho}\mathbf{u}- \hat{\mathbf{H}} \hat{\mathbf{h}}_d - \omega \mathbf{r} \right) \right\} \\
&- \mathbf{v}^H\mathbf{U}\bm{\Sigma}^+ \mathbf{U}^H \mathbf{v}\\
\textrm{s.t.}\;  &\|\mathbf{v}\|^2 \leq N,
\end{aligned}
\end{equation}
where we have used $\left[ \bar{v}_{11} \; \mathbf{r}^H ; \mathbf{r} \; \mathbf{R}  \right] = \bar{\mathbf{V}}$ and the eigen-decomposition $ \hat{\mathbf{H}} \hat{\mathbf{H}}^H + \omega \mathbf{R} = \mathbf{U}(\bm{\Sigma}^+ + \bm{\Sigma}^-)\mathbf{U}^H$.  Since the constraint of problem \eqref{v_subproblem} is convex and its objective function can be expressed as a difference of two convex functions when $\mathbf{u}$ is fixed, we can apply the BSUM method to solve it approximately. Note that $\mathbf{v}^H \mathbf{U} \bm{\Sigma}^-\mathbf{U}^H  \mathbf{v}$ contains the concave component of the original objective function in \eqref{weighted_sum_problem}, and by only approximating the convex component, i.e., $\mathbf{v}^H \mathbf{U} \bm{\Sigma}^+\mathbf{U}^H  \mathbf{v} $, we can reduce the approximation error and potentially achieve a faster convergence speed. Specifically, by resorting to the first-order Taylor expansion at a given point $\bar{\mathbf{v}}$, problem \eqref{v_subproblem} can be approximated by
\begin{equation} \label{v_subproblem_appro}
\begin{aligned}
\min\limits_{\mathbf{v}} \;& \mathbf{v}^H(\mathbf{I} - 2 \rho\mathbf{U} \bm{\Sigma}^-\mathbf{U}^H) \mathbf{v}  \\
&+ 2  \Re\left\{\mathbf{v}^H \left(\rho \bm{\lambda}-\mathbf{u}- 2\rho \hat{\mathbf{H}} \hat{\mathbf{h}}_d -2\rho  \omega \mathbf{r} \right) \right\} \\
& - 4 \rho \Re\{ (\mathbf{U}\bm{\Sigma}^+ \mathbf{U}^H \bar{\mathbf{v}})^H (\mathbf{v} - \bar{\mathbf{v}}) \}\\
\textrm{s.t.}\;  &\|\mathbf{v}\|^2 \leq N. 
\end{aligned}
\end{equation}
 Next, by resorting to the first-order optimality condition of problem \eqref{v_subproblem_appro}, we have
\begin{equation} \label{v_step1}
\mathbf{v}(\mu) = \left((1+\mu)\mathbf{I} - 2 \rho \mathbf{U} \bm{\Sigma}^-\mathbf{U}^H\right)^{-1} \mathbf{b},
\end{equation}
where $\mathbf{b} = 2\rho \mathbf{U}\bm{\Sigma}^+ \mathbf{U}^H \bar{\mathbf{v}}   -\rho \bm{\lambda}+\mathbf{u}+ 2\rho \hat{\mathbf{H}} \hat{\mathbf{h}}_d +2\rho \omega \mathbf{r}$ and $\mu$ denotes the dual variable associated with the constraint $\|\mathbf{v}\|^2 \leq N$. If $\| \mathbf{v}(0)\|^2 \leq N$, then $\mathbf{v}(0)$ is the optimal solution of problem \eqref{v_subproblem_appro}; otherwise, the optimal dual variable $\mu$ can be found via bisection search. In the special case of $\bm{\Sigma}^{-} = \mathbf{0}$, problem \eqref{v_subproblem_appro} admits a closed-form solution which is given by \cite{zhao2019intelligent}
\begin{equation} \label{v_step2}
\mathbf{v} = \left\{ \begin{array}{l}
\mathbf{b} ,\;\textrm{if}\; \|\mathbf{b} \|^2 \leq N,\\
\frac{\mathbf{b}}{ \sqrt{\|\mathbf{b}\|^2/N}},\;\textrm{otherwise}.
\end{array}
\right.
\end{equation}

In the $\mathbf{u}$-optimization step, we have
\begin{equation} \label{u_subproblem}
\begin{aligned}
\min\limits_{\mathbf{u}} \; & \|\mathbf{v}-\mathbf{u}+\rho \bm{\lambda}\|^2 \\
\textrm{s.t.}\; & u_n \in \mathcal{F}_d,\;\forall n \in\mathcal{N},
\end{aligned}
\end{equation}
which can be optimally solved in parallel by mapping each element of $\mathbf{v}+\rho \bm{\lambda}$ to the nearest discrete value in $\mathcal{F}_d$.

Besides, in the outer loop, the dual variable $\bm{\lambda}$ is updated by
\begin{equation}
\bm{\lambda} = \bm{\lambda} + \frac{1}{\rho}(\mathbf{v} - \mathbf{u}).
\end{equation}
To summarize, we can one-dimensionally search over $\omega$ and for each fixed $\omega$, we employ the PDD-based algorithm to solve problem \eqref{weighted_sum_problem} and obtain the candidate IRS phase-shift vector $\mathbf{v}[\omega]$; then with fixed $\mathbf{v}[\omega]$ and the MRT-based solution for $\mathbf{w}[\omega]$, the corresponding AP transmit power $p[\omega]$ can be easily found via bisection search; finally, we choose the best solution pair $(\mathbf{v}[\omega], \mathbf{w}[\omega])$ that achieves the lowest transmit power. The detailed steps of the proposed WSMax algorithm to solve problem \eqref{single_user_problem} are given in Algorithm \ref{algorithm_WSMax}. It is noteworthy that the PDD-based algorithm in Algorithm \ref{algorithm_WSMax} is guaranteed to converge to the set of stationary solutions of problem \eqref{weighted_sum_problem} in the continuous phase-shift case and it can also achieve good performance in the discrete phase-shift case \cite{zhao2019intelligent}. While for the overall Algorithm \ref{algorithm_WSMax}, it is generally difficult to analyze its convergence property since the outage probability $\mathcal{C} (\mathbf{v};p) $ does not have an analytic expression and there exist discrete variables in problem \eqref{single_user_problem}. However, Algorithm \ref{algorithm_WSMax} is able to converge to a high-quality suboptimal solution, as will be verified by numerical simulations later in Section \ref{section_simulation}.

\begin{algorithm}[t]
	\caption{Proposed WSMax algorithm for Solving Problem \eqref{single_user_problem}} \label{algorithm_WSMax}
	\begin{algorithmic}[1]
		\STATE  \textbf{Input}: $p^{\delta}$ and $\epsilon^{\delta}$. Initialize $\omega_l$, $\omega_u$ and $\Delta \omega$, let $\omega = \omega_l$ and $p^{\textrm{best}} = \infty$.
		\REPEAT
		
		\STATE Apply the PDD-based algorithm in \cite{zhao2019intelligent} to solve problem \eqref{weighted_sum_problem} with the $\mathbf{v}$-optimization step replaced by \eqref{v_step1} or \eqref{v_step2} and obtain $\mathbf{v}[\omega]$.
			\STATE Let $p^l = 0$ and set $p^u$ to a large number such that $\mathcal{C} (\mathbf{v}[\omega];p^u) < \epsilon$.
		\REPEAT 
		\STATE Let $p[\omega] = (p^l+p^u)/2$ and calculate $\mathcal{C} (\mathbf{v}[\omega];p[\omega] )$.
		\STATE \textbf{If} {$\mathcal{C} (\mathbf{v}[\omega];p[\omega] )<\epsilon$}, \textbf{then} $p^u = p[\omega]$, \textbf{else}  $p^l = p[\omega]$, \textbf{end}.
		\UNTIL{$p^u-p^l \leq p^{\delta}$ and $|\mathcal{C} (\mathbf{v}[\omega];p[\omega] )-\epsilon| < \epsilon^{\delta}$.}
		\STATE Obtain $\mathbf{w}[\omega] = \sqrt{p[\omega] }\frac{( \mathbf{v}[\omega]^H \hat{\mathbf{H}} + \hat{\mathbf{h}}_d^H)^H}{\| \mathbf{v}[\omega]^H \hat{\mathbf{H}} + \hat{\mathbf{h}}_d^H\|}$.
		\IF {$p[\omega] < p^{\textrm{best}}$}
		\STATE $(\mathbf{v}^{\textrm{best}}, \mathbf{w}^{\textrm{best}}) \leftarrow (\mathbf{v}[\omega], \mathbf{w}[\omega])$, $p^{\textrm{best}} \leftarrow p[\omega]$.
		\ENDIF
		\STATE $\omega \leftarrow \omega + \Delta \omega$.
		\UNTIL{$\omega> \omega_u$.}
		\STATE \textbf{Output}: $\mathbf{v}^{\textrm{best}}$ and  $\mathbf{w}^{\textrm{best}}$.
	\end{algorithmic}
\end{algorithm}

\begin{remark} \label{remark1}
\emph{Note that problem \eqref{weighted_sum_problem} needs to be solved multiple times (each with a different value of $\omega$) in the proposed WSMax algorithm, and the number of times is determined by the granularity during the one-dimensional search for $\omega$. Through numerical simulations, we find that only very coarse search over $\omega$ is required to achieve near-optimal performance, thus this one-dimensional search will not lead to unaffordable computational complexity.}
\end{remark}

\begin{remark}
\emph{There are two alternative ways to solve problem \eqref{v_subproblem_eq2}, one is to find all Pareto-boundary points of the MVR-variance region (since we want to maximize the MVR and variance simultaneously) and employ a rate-profile-type approach \cite{Zhang2010rateprofile}; while the other is to formulate a variance-constrained MSP maximization problem and search the variance region piece-by-piece. However, for the former approach, solving the resulting problem is highly involved with complicated MVR and variance constraints, while for the latter approach, the variance-constrained MSP maximization problem is also very difficult to solve since the variance region can be very sparse (as shown in Fig. \ref{fig:region_discrete}); therefore we adopt the more tractable WSMax approach in this paper.}
\end{remark}

\subsection{Relationship with Baseline Algorithms} \label{Section_heuristic}
In this subsection, we introduce three baseline algorithms to solve problem \eqref{v_subproblem_eq2} and show that they serve as special cases of the proposed WSMax algorithm.

\subsubsection{MVR Maximization}
The first baseline algorithm aims to maximize the MVR only, which leads to the following problem:
\begin{equation} \label{MVR_maximization_problem}
\begin{aligned}
\max \limits_{\mathbf{v}} \; & \frac{\mathbf{v}^H \hat{\mathbf{H}} \hat{\mathbf{H}}^H \mathbf{v}+2\Re\{\mathbf{v}^H \hat{\mathbf{H}} \hat{\mathbf{h}}_d \}+\hat{\mathbf{h}}_d^H \hat{\mathbf{h}}_d   }{ \mathbf{v}^H \mathbf{R} \mathbf{v}+2\Re\{ \mathbf{v}^H\mathbf{r}\}+\bar{v}_{11} } \\
\textrm{s.t.}\; & v_n \in \mathcal{F}_d,\;\forall n \in \mathcal{N}.
\end{aligned}
\end{equation}
Problem \eqref{MVR_maximization_problem} is similar to the achievable rate maximization problem in \cite{Zhao2020intelligent}, therefore the penalized Dinkelbach-BSUM algorithm therein can be applied to efficiently solve it, and the details are omitted for brevity. Note that problem \eqref{MVR_maximization_problem} can be interpreted as a special instance of problem \eqref{weighted_sum_problem} since when $\omega$ is negative, $-\omega$ can be viewed as the Dinkelbach parameter when applying the penalized Dinkelbach-BSUM algorithm. After obtaining the optimized $\mathbf{v}$, we can employ a similar bisection search over $p$ as that in Algorithm \ref{algorithm_WSMax} to find the required transmit power.

\subsubsection{MPV Maximization}
In the second baseline algorithm, we aim to minimize the lower bound of the outage probability. Specifically, from the Markov's inequality, we have that $\textrm{Pr}(x \geq t)  \leq t^{-1}\mathbb{E}\{x\}$ holds for any non-negative random variable $x$ \cite{Ntranos2009outage}, thus we can obtain\footnote{It is difficult to obtain a sensible upper bound for the outage probability, therefore we resort to its lower bound instead.}
\begin{equation} \label{Markov_inequality}
\begin{aligned}
& \textrm{Pr}\left( {|( \mathbf{v}^H \mathbf{H} + \mathbf{h}_{d}^H) \mathbf{w}|^2} < \sigma^2 \eta \right) \\
& = 1- \textrm{Pr}\left( {|( \mathbf{v}^H \mathbf{H} + \mathbf{h}_{d}^H) \mathbf{w}|^2} \geq \sigma^2 \eta \right) \\
& \geq 1- \frac{1}{\eta \sigma^2}\mathbb{E} \left\{ |( \mathbf{v}^H \mathbf{H} + \mathbf{h}_{d}^H) \mathbf{w}|^2\right\}.
\end{aligned}
\end{equation}
It can be observed that minimizing this lower bound is equivalent to maximizing the MPV $\mathbb{E} \left\{ |( \mathbf{v}^H \mathbf{H} + \mathbf{h}_{d}^H) \mathbf{w}|^2\right\}$. By resorting to \cite[Proposition 1]{Zhao2020intelligent} and employing the optimal MRT-based solution of $\mathbf{w}$, we have the following approximated problem of \eqref{outage_minimize} (i.e., the MPV maximization problem):
\begin{equation} \label{outage_maximize_appro_eq}
\begin{aligned}
\max \limits_{\mathbf{\mathbf{v}} } \; & \mathbf{v}^H (\hat{\mathbf{H}}\hat{\mathbf{H}}^H + \mathbf{R} ) \mathbf{v} + 2\Re \{\mathbf{v}^H (\hat{\mathbf{H}}\hat{\mathbf{h}}_d + \mathbf{r} ) \}\\
\textrm{s.t.}\;  & v_n \in \mathcal{F}_d,\;\forall n \in \mathcal{N},
\end{aligned}
\end{equation}
which is a special instance of problem \eqref{weighted_sum_problem} when $\omega=1$.

\subsubsection{MSP Maximization} Thirdly, based on the intuition that the MSP should be as large as possible, we ignore the variance term $\tilde{\mathbf{v}}^H \bar{\mathbf{V}} \tilde{\mathbf{v}}$ and consider the following MSP maximization problem directly:
\begin{equation} \label{MSP_maximization}
\begin{aligned}
\max \limits_{\mathbf{\mathbf{v}} } \; &  \mathbf{v}^H \hat{\mathbf{H}}\hat{\mathbf{H}}^H \mathbf{v} + 2\Re \{\mathbf{v}^H \hat{\mathbf{H}}\hat{\mathbf{h}}_d \} \\
\textrm{s.t.}\;  & v_n \in \mathcal{F}_d,\;\forall n \in \mathcal{N}.
\end{aligned}
\end{equation}
Note that problem \eqref{weighted_sum_problem} reduces to problem \eqref{MSP_maximization} when $\omega=0$ and the latter is suitable for the case when the CSI errors are uncorrelated (corresponding to the continuous phase-shift case and $N_r=N+1$).\footnote{In this case, maximizing the MSP is optimal since $ \bar{\mathbf{V}}$ becomes a diagonal matrix and the variance $\tilde{\mathbf{v}}^H  \bar{\mathbf{V}} \tilde{\mathbf{v}} $ becomes a constant regardless of the design of the reflection phase-shift vector $\mathbf{v}$.}

Therefore, compared to problems \eqref{MVR_maximization_problem}, \eqref{outage_maximize_appro_eq} and \eqref{MSP_maximization}, the formulation of problem \eqref{weighted_sum_problem} is more general and by one-dimensionally searching over $\omega$ and efficiently solving \eqref{weighted_sum_problem}, we are expected to achieve better performance than solving \eqref{MVR_maximization_problem}, \eqref{outage_maximize_appro_eq} or \eqref{MSP_maximization} individually in general.

\subsection{Complexity Analysis}
The complexity of the proposed WSMax algorithm (i.e., Algorithm \ref{algorithm_WSMax}) is mainly due to the PDD-based algorithm used to solve problem \eqref{weighted_sum_problem}, thus can be shown to be of order $\mathcal{O}(N_{\omega} I_oI_iN^3$ $ \log(1/\epsilon_{\textrm{bi}}))$, where $N_{\omega}$ denotes the number of $\omega$-values traversed during one-dimensional search, $I_o$ and $I_i$ denote the maximum outer and inner iteration numbers of the PDD-based algorithm, and $\epsilon_{\textrm{bi}}$ represents the accuracy of the bisection method over the dual variable $\mu$ in \eqref{v_step1}. Besides, the complexity of the MVR maximization algorithm is $\mathcal{O}(I_{\textrm{BCD}}2^{Q}N^3 + I_P I_{D}N^{3}\log(1/\epsilon_{\textrm{bi}}) )$ \cite{Zhao2020intelligent}, where $I_{\textrm{BCD}}$, $I_P$ and $I_D$ denote the number of iterations required by the block coordinate descent (BCD) method, penalty method and Dinkelbach-BSUM method therein, respectively. For the MPV and MSP maximization algorithms, the complexity is given by $\mathcal{O}(I_oI_iN^2)$ \cite{zhao2019intelligent}. 

\section{Multiuser System} \label{Section_multiuser}
In this section, we consider the general multiuser case, where multiple users are assumed to share the same time-frequency resource and there exists multiuser interference in general. Since it is difficult to characterize the distributions of the users' SINRs, the WSMax algorithm designed for the single-user system does not apply in this case. Thus, we propose a two-stage CSSCA algorithm to solve problem \eqref{outage_problem}. Specifically, we first transform problem \eqref{outage_problem} into a more tractable form by utilizing smooth approximation and then construct surrogate functions for the outage probabilities using randomly generated CSI error samples based on their known statistics. Next, in the first stage, the discrete IRS phase shifts are relaxed into continuous ones and are jointly optimized with the active precoders at the AP. Then, in the second stage, by quantizing the continuous phase shifts to discrete values and keeping them fixed, the active precoders are further optimized to compensate for the outage performance loss caused by phase quantization.

\subsection{Problem Transformation}
First, let $i$ denote the time slot index, we have $\textrm{Pr} (\textrm{SINR}_k \leq \eta_k )= \lim_{n\rightarrow \infty} \frac{\sum\nolimits_{i=1}^n u(\eta_k - \textrm{SINR}_k[i])}{n} =\mathbb{E}\{ u(\eta_k - \textrm{SINR}_k) \}$ by definition, where $u (\cdot)$ is the step function, i.e., the SINR outage probability can be interpreted as the expectation of a step function parameterized by $\eta_k - \textrm{SINR}_k$. Then, to resolve the difficulty brought by the non-smoothness of the step function, we resort to the following smooth approximate function:
\begin{equation} \label{approximate_step_function}
\hat{u}_{\vartheta}(x) = \frac{1}{1+e^{-\vartheta x}},
\end{equation}
where the smooth parameter $\vartheta$ is used to control the approximation error (i.e., larger $\vartheta$ leads to less approximation error). By replacing the step function $u (\cdot)$ with its smooth approximation $\hat{u}_{\vartheta}(\cdot)$, we can obtain an approximation of problem \eqref{outage_problem} as follows:
\begin{equation} \label{outage_problem_equi2}
\begin{aligned}
\min\limits_{\{\mathbf{w}_k\},\mathbf{v}} \; & \sum\limits_{k \in\mathcal{K}} \| \mathbf{w}_k\|^2\\
\textrm{s.t.}\; & \mathbb{E} \{ \hat{u}_{\vartheta}( q_k( \{\mathbf{w}_k\},\mathbf{v}; \mathcal{H}) ) \} \leq \epsilon_k, \;\forall k \in \mathcal{K},\\
&  v_n \in \mathcal{F}_d,\;\forall n \in \mathcal{N},
\end{aligned}
\end{equation}
where $q_k(\{ \mathbf{w}_k\} ,\mathbf{v}; \mathcal{H}) \triangleq \eta_k (\sum\nolimits_{j\in\mathcal{K}\backslash k}|(\mathbf{v}^H \mathbf{H}_{k}  + \mathbf{h}_{d,k}^H)\mathbf{w}_j|^2 $ $+\sigma_k^2)  - {|(\mathbf{v}^H \mathbf{H}_{k} + \mathbf{h}_{d,k}^H) \mathbf{w}_k|^2}$ and $\mathcal{H} \triangleq  \{\mathbf{H}_{k},\mathbf{h}_{d,k} \}_{k \in \mathcal{K}}$.

Problem \eqref{outage_problem_equi2} is a non-convex constrained stochastic optimization problem with the random state given by the CSI errors $\bm{\xi}\triangleq \{\Delta \tilde{\mathbf{H}}_{k}\}_{k\in\mathcal{K}}$. Define $\bm{\varpi} \triangleq [\mathbf{w}_1^T,\cdots,\mathbf{w}_K^T,\mathbf{v}^T]^T$ as the composite optimization variable, and let $z_k(\bm{\varpi};\mathcal{H}) = \eta_k (\sum\nolimits_{j\in\mathcal{K}\backslash k}|(\bm{\varpi}^H \mathbf{A}^H\mathbf{H}_{k}  + \mathbf{h}_{d,k}^H)\mathbf{B}_j\bm{\varpi}|^2+\sigma_k^2) 
- {|(\bm{\varpi}^H \mathbf{A}^H \mathbf{H}_{k} + \mathbf{h}_{d,k}^H) \mathbf{B}_k\bm{\varpi}|^2}$ and $g_k( \bm{\varpi };\mathcal{H}) \triangleq \hat{u}_{\vartheta}\left( q_k(\{ \mathbf{w}_k\} ,\mathbf{v}; \mathcal{H}) \right)
=\hat{u}_{\vartheta} (z_k(\bm{\varpi};\mathcal{H}) )$, where $\mathbf{A} \in \{ 0,1\}^{N \times (K M +N)}$ and $\{\mathbf{B}_k \in \{0,1 \}^{M \times (K M+N)} \}$ are selection matrices that satisfy $\mathbf{A} \bm{\varpi} = \mathbf{v}$ and $\mathbf{B}_k \bm{\varpi} = \mathbf{w}_k$, respectively.\footnote{Note that such selection matrices always exist.} 
Then, by replacing $\mathbf{v}$ and $\{\mathbf{w}_k \}$ with $\mathbf{A} \bm{\varpi}$ and $\{\mathbf{B}_k \bm{\varpi}\}$, respectively, problem \eqref{outage_problem_equi2} can be rewritten into a more compact form as\footnote{In the following, the notations $q(\bm{\varpi};\mathcal{H})$ and $q(\bm{\varpi};\bm{\xi})$ will be used interchangeably to denote that a certain function $q(\cdot;\cdot)$ depends on the optimization variable $\bm{\varpi}$ and random state $\bm{\xi}$. }
\begin{subequations} \label{outage_problem_equi3}
\begin{align}
\min\limits_{ \bm{\varpi }} \; & \sum\limits_{k \in \mathcal{K}} \bm{\varpi}^H \mathbf{B}_k \mathbf{B}_k \bm{\varpi} \\
\textrm{s.t.}\; & f_k(\bm{\varpi }) \triangleq \mathbb{E}\left\{ g_k( \bm{\varpi };\mathcal{H}) \right\} \leq \epsilon_k, \;\forall k \in \mathcal{K},\\
& v_n \in \mathcal{F}_d,\;\forall n \in \mathcal{N}. \label{discrete_constraints}
\end{align}
\end{subequations}
Note that although the formulation of problem \eqref{outage_problem_equi3} is similar to that considered in \cite{zhou2020framework}, the Bernstein-type approach employed in \cite{zhou2020framework} is not directly applicable here due to the discrete IRS phase-shift constraints. Besides, the proposed two-stage CSSCA algorithm works for any channel estimation error distribution, while the Bernstein-type approach only works for Gaussian error distributions.

\subsection{Proposed Two-Stage CSSCA Algorithm}
In this subsection, we leverage the stochastic optimization framework in \cite{liu2018stochastic} and propose a novel two-stage CSSCA algorithm to address problem \eqref{outage_problem_equi3}. In the first stage, in order to make problem \eqref{outage_problem_equi3} tractable, we relax the discrete constraints in \eqref{discrete_constraints} to $|v_n|^2 \leq 1$.\footnote{We allow the reflection amplitudes to be in the interval $[0, 1]$, which has been shown in \cite{zhao2019intelligent} to help accelerate the convergence. After obtaining the optimized IRS phase shifts $\mathbf{v}$, we can project each of its entries independently onto $\mathcal{F}_{d}$ to obtain a unit-modulus feasible solution in case $|v_n|<1, \exists n$. Besides, although this relaxation may not be tight theoretically, it works well in all of our simulations and the converged solution almost always satisfies $|v_n|=1,\forall n \in \mathcal{N}$.} Then, we have the following problem:
\begin{equation} \label{step1_problem}
\begin{aligned}
\min\limits_{ \bm{\varpi }} \; & \sum\limits_{k \in \mathcal{K}} \bm{\varpi}^H \mathbf{B}_k \mathbf{B}_k \bm{\varpi} \\
\textrm{s.t.}\; & f_k(\bm{\varpi })  \leq \epsilon_k, \;\forall k \in \mathcal{K},\\
 & |v_n|^2  \leq 1,\;\forall n \in \mathcal{N}.
\end{aligned}
\end{equation}
After solving problem \eqref{step1_problem} and obtaining the optimized IRS reflection coefficient vector $\mathbf{v}^o$, we project their phase shifts into $\mathcal{F}_d$ to obtain the quantized $\mathbf{v}^q$. Then, in the second stage with fixed $\mathbf{v}^q$, we solve the following problem:
\begin{equation} \label{step2_problem}
\begin{aligned}
\min\limits_{ \mathbf{w}} \;&  \| \mathbf{w}\|^2 \\
\textrm{s.t.}\; & \tilde{f}_k( \mathbf{w}) \leq \epsilon_k, \;\forall k \in \mathcal{K},
\end{aligned}
\end{equation}
where $\mathbf{w} \triangleq [\mathbf{w}_1^T,\cdots, \mathbf{w}_K^T]^T$, $\tilde{f}_k( \mathbf{w}) = \mathbb{E}\{ \tilde{g}_k(\mathbf{w};{\mathbf{h}}) \}$, $\mathbf{h} \triangleq  \{\mathbf{h}_k =\mathbf{H}_{k}^H \mathbf{v}^q + {\mathbf{h}}_{k}\}$ and $\tilde{g}_k(\mathbf{w};\mathbf{h}) = \hat{u}_{\vartheta}(\eta_k ({\sum\nolimits_{j\in\mathcal{K}\backslash k}|\mathbf{h}_k^H \mathbf{w}_j|^2+\sigma_k^2}) - {|\mathbf{h}_k \mathbf{w}_k|^2})$. Note that problem \eqref{step2_problem} reduces to the conventional robust beamforming problem for a multiuser MISO downlink system, and it can be viewed as a special case of problem \eqref{step1_problem} with fixed $\mathbf{v} = \mathbf{v}^q$. 

In the following, we present the proposed CSSCA algorithm to iteratively solve problem \eqref{step1_problem} in the first stage. Specifically, in the $t$-th iteration, convex surrogate functions $\{ \bar{f}^t_k(\bm{\varpi })\}$ are constructed to deal with the unavailability of closed-form expressions for the approximated outage probabilities $\{ f_k(\bm{\varpi }) \}_{k \in \mathcal{K}}$, which can be expressed as \cite{liu2018stochastic}
\begin{equation} \label{structured_s_function_simplified}
\bar{f}^t_k(\bm{\varpi }) = f_k^t + 2 \Re \left\{ (\mathbf{f}_k^{t})^H (\bm{\varpi} - \bm{\varpi}^t) \right\} + \tau_k \| \bm{\varpi} - \bm{\varpi}^t\|^2,
\end{equation}
where $\tau_k > 0$ a positive constant and the term $\tau_k \| \bm{\varpi} - \bm{\varpi}^t\|^2$ is added to ensure strong convexity of $\bar{f}^t_k(\bm{\varpi })$, $f_k^t$ is an approximation for $\mathbb{E}\{g_k(\bm{\varpi}^t; \bm{\xi}) \} $ and $\mathbf{f}_k^t$ is an approximation for the conjugate gradient $\nabla_{\bm{\varpi}^*} \mathbb{E} \{ g_k(\bm{\varpi}^t; \bm{\xi})\}$. $f_k^t$ and $\mathbf{f}_k^t$ are iteratively updated according to
\begin{equation} \label{f_constant_update_new}
f_k^t = \frac{1}{L} \sum\limits_{l=1}^{L} g_k(\bm{\varpi}^t; \bm{\xi}^l),
\end{equation}
\begin{equation} \label{f_gradient_update_new}
\mathbf{f}_k^t = (1-\rho^t) \mathbf{f}_k^{t-1} + \rho^t \frac{1}{T_H}\sum\limits_{l=1}^{T_H} \nabla_{\bm{\varpi}^*} g_k(\bm{\varpi}^t; \bm{\xi}^l),
\end{equation}
where $f_k^{-1} =  0$, $\mathbf{f}_k^{-1} = \mathbf{0}$, $L$ and $T_H$ denote the numbers of channel error samples used to approximate $\mathbb{E}\{g_k(\bm{\varpi}^t; \bm{\xi}) \} $  and $\nabla_{\bm{\varpi}^*} \mathbb{E} \{ g_k(\bm{\varpi}^t; \bm{\xi})\}$, respectively, $\rho^t \in (0, 1]$ is a sequence properly chosen according to Assumption 5 in \cite{liu2018stochastic}.  
Note that in \eqref{f_constant_update_new} and \eqref{f_gradient_update_new}, multiple channel error samples are generated to improve the approximations $f_k^t$ and $\mathbf{f}_k^t$ in each iteration, which can help accelerate the speed of the proposed algorithm to converge to the feasible region of problem \eqref{step1_problem}. Besides, as $t \rightarrow \infty$, the following asymptotic consistency properties of the surrogate functions are satisfied \cite{liu2018stochastic}: $\lim_{t \rightarrow \infty} |\bar{f}^t_k(\bm{\varpi}^t)  - f_k(\bm{\varpi}^t) |  = 0$, $\lim_{t \rightarrow \infty} \| \nabla_{\bm{\varpi}^*}\bar{f}^t_k(\bm{\varpi}^t) -   \nabla_{\bm{\varpi}^*} f_k(\bm{\varpi}^t) \| = 0$, $\forall k \in \mathcal{K}$. This means that the approximations $\bar{f}^t_k(\bm{\varpi}^t)$ and $\nabla_{\bm{\varpi}^*}\bar{f}^t_k(\bm{\varpi}^t)$ can converge to the true values of $f_k(\bm{\varpi}^t)$ and its conjugate gradient with respect to $\bm{\varpi}$, respectively, which is essential for guaranteeing the convergence of the proposed algorithm. 
For each channel error sample, the conjugate gradient $\nabla_{\bm{\varpi}^*} g_k(\bm{\varpi}^t; \bm{\xi}^l)$ is obtained by (ignoring the iteration index $t$ and sample index $l$ for simplicity and applying the chain rule)
\begin{equation} \label{CSSCA_gradient}
\begin{aligned}
\nabla_{\bm{\varpi}^*} g_k (\bm{\varpi}; \mathcal{H}) & =  \hat{u}_{\vartheta}'(z_k(\bm{\varpi};\mathcal{H}))  \mathbf{z}_k'(\bm{\varpi};\mathcal{H}) \\
& = \frac{\vartheta e^{-\vartheta z_k(\bm{\varpi};\mathcal{H})}}{\left(1+e^{-\vartheta z_k(\bm{\varpi};\mathcal{H})}\right)^2} \mathbf{z}_k'(\bm{\varpi};\mathcal{H}),
\end{aligned}
\end{equation}
where $\mathbf{z}_k'(\bm{\varpi};\mathcal{H}) = 2\eta_k ( \sum\nolimits_{j \in \mathcal{K}\backslash k} \mathbf{q}_{kj}(\bm{\varpi}; \mathcal{H} ) ) - 2 \mathbf{q}_{kk}(\bm{\varpi}; \mathcal{H})$ and $\mathbf{q}_{kj}(\bm{\varpi}; \mathcal{H}) =  \mathbf{A}^H \mathbf{H}_k \mathbf{B}_j \bm{\varpi} \bm{\varpi}^H\mathbf{B}_j^H $ $ \mathbf{H}_k^H \mathbf{A}  \bm{\varpi} 
+ \bm{\varpi}^H \mathbf{A}^H $ $   \mathbf{H}_{k}  \mathbf{B}_j   \bm{\varpi}  \mathbf{B}_j^H \mathbf{H}_k^H  \mathbf{A}  \bm{\varpi}  + \mathbf{B}_j^H \mathbf{h}_{d,k} \mathbf{h}_{d,k}^H \mathbf{B}_j \bm{\varpi} 
+ \mathbf{A}^H \mathbf{H}_k \mathbf{B}_j   \bm{\varpi}  \bm{\varpi}^H $ $ \mathbf{B}_j^H \mathbf{h}_{d,k} + \bm{\varpi}^H\mathbf{A}^H \mathbf{H}_k  $ $ \mathbf{B}_j \bm{\varpi} \mathbf{B}_j^H \mathbf{h}_{d,k} 
+ \mathbf{h}_{d,k}^H \mathbf{B}_j \bm{\varpi} \mathbf{B}_j^H \mathbf{H}_k^H \mathbf{A} \bm{\varpi}$.

Then, we solve the following problem in the $t$-th iteration:
\begin{equation} \label{outage_sub1}
\begin{aligned}
\bar{\bm{\varpi }}^t =  \arg\min\limits_{ \bm{\varpi }} \;  & \sum\limits_{k \in \mathcal{K}} \bm{\varpi}^H \mathbf{B}_k \mathbf{B}_k \bm{\varpi} \\
 \textrm{s.t.}\; & \bar{f}^t_k(\bm{\varpi }) - \epsilon_k \leq 0, \;\forall k \in \mathcal{K},\\
 & |v_n|^2 \leq 1,\;\forall n \in \mathcal{N},
\end{aligned}
\end{equation}
which can be further expressed as a convex second-order cone program (SOCP) problem and efficiently solved by off-the-shelf solvers, such as CVX \cite{CVX}. If problem \eqref{outage_sub1} is infeasible, we solve the following problem instead:
\begin{equation} \label{outage_sub2}
\begin{aligned}
\bar{\bm{\varpi }}^t = \arg\min\limits_{ \bm{\varpi },\;\alpha} \; & \alpha\\
\textrm{s.t.}\; &  \bar{f}^t_k(\bm{\varpi }) - \epsilon_k \leq \alpha, \;\forall k \in \mathcal{K},\\
& |v_n|^2 \leq 1,\;\forall n \in \mathcal{N},
\end{aligned}
\end{equation}
which minimizes the gap between the surrogate functions $\{\bar{f}^t_k(\bm{\varpi })\}_{k\in\mathcal{K}}$ and the corresponding outage probability targets $\{\epsilon_k\}_{k\in\mathcal{K}}$, i.e., $\alpha$. Solving problem \eqref{outage_sub2} helps to pull the solution to the feasible region of problem \eqref{step1_problem} when the current problem at iteration $t$ is infeasible.
Given $\bar{\bm{\varpi }}^t$ in one of the above two cases, $\bm{\varpi}$ is updated according to
\begin{equation} \label{w_update}
\bm{\varpi }^{t+1} = (1-\gamma^t) \bm{\varpi }^{t} + \gamma^t \bar{\bm{\varpi }}^t,
\end{equation}
where $\{\gamma^t \}$ is a sequence satisfying $\gamma^t \rightarrow 0$, $\sum_t \gamma^t = \infty$ and $\sum_t (\gamma^t)^2 < \infty$. 

In the second stage, problem \eqref{step2_problem} is iteratively solved by applying a similar CSSCA algorithm with fixed $\mathbf{v}^q$. To summarize, the proposed two-stage CSSCA algorithm to solve problem \eqref{outage_problem} is listed in Algorithm \ref{algorithm_CSSCA}. Besides, according to \cite[Theorem 1]{liu2018stochastic}, the CSSCA algorithm in both stages can converge to the set of stationary solutions of problems \eqref{step1_problem} and \eqref{step2_problem}, respectively, almost surely, therefore the convergence of the overall Algorithm \ref{algorithm_CSSCA} can be guaranteed.
\begin{algorithm}[!h]
	\caption{Proposed Two-Stage CSSCA Algorithm for Solving Problem \eqref{outage_problem}} \label{algorithm_CSSCA}
	\begin{algorithmic}[1]
		\STATE \textbf{Input}: $\{\rho^t \}$, $\{\gamma^t \}$, $L$, $T_H$ and $\xi_o$. \textbf{Initialize}: $\bm{\varpi}^0$. Set $t=0$. 
		\STATE \textbf{Stage I}:
		\STATE Generate $\{\bm{\xi}^l\}$ according to $\{\frac{1}{\sqrt{p_{u,k}} }  (\mathbf{V}^{\dagger})^H \mathbf{N}_{u,k}^H \} $. Update the surrogate functions $\bar{f}^t_k(\bm{\varpi }),\forall k $ according to \eqref{structured_s_function_simplified}, using $\{\bm{\xi}^l\}$ and $\bm{\varpi}^t$.
		\STATE {\textbf{If} Problem \eqref{outage_sub1} is feasible}, \textbf{then} solve problem \eqref{outage_sub1} to obtain $\bar{\bm{\varpi }}^t$, \textbf{else} solve problem \eqref{outage_sub2} to obtain $\bar{\bm{\varpi }}^t$, \textbf{end}.
		\STATE Update $\bm{\varpi }^{t+1}$ by \eqref{w_update}.
		\STATE Let $t = t + 1$, if the fractional decrease of $f_k^t$ is larger than the threshold $\xi_o$, return to \textbf{Step 3}, otherwise, go to \textbf{Step 7}.
		\STATE Extract the optimized IRS reflection coefficients $\mathbf{v}^o$ from $\bm{\varpi}$ and quantize $\mathbf{v}^o$ to $\mathbf{v}^q$.
		\STATE \textbf{Stage II}: Repeat \textbf{Steps 3-6} with fixed $\mathbf{v}^q$ to obtain optimized $\{\mathbf{w}_k\}$.
		\STATE \textbf{Output}: $\mathbf{v}^{q}$ and $\{\mathbf{w}_k\}$.
	\end{algorithmic}
\end{algorithm}

\begin{remark}
	\emph{Since we use the smooth approximation function in \eqref{approximate_step_function} to imitate the behavior of the step function, the gradient $\nabla_{\bm{\varpi}^*} g_k (\bm{\varpi}; \mathcal{H}) $ may approach to zero due to the term $\frac{\vartheta e^{-\vartheta z_k(\bm{\varpi};\mathcal{H})}}{\left(1+e^{-\vartheta z_k(\bm{\varpi};\mathcal{H})}\right)^2}$ in \eqref{CSSCA_gradient} when $|z_k(\bm{\varpi};\mathcal{H})|$ is too large. This causes the so-called ``vanishing gradient'' problem that would prevent the proposed algorithm from updating the variable $\bm{\varpi}$. To tackle this difficulty, we modify the gradient in \eqref{CSSCA_gradient} as follows:
	\begin{equation}
	\nabla_{\bm{\varpi}^*} g_k (\bm{\varpi}; \mathcal{H}) = \frac{\vartheta e^{-\bar{z}_k(\bm{\varpi};\mathcal{H})}}{\left(1+e^{-\bar{z}_k(\bm{\varpi};\mathcal{H})}\right)^2}\mathbf{z}_k'(\bm{\varpi};\mathcal{H}),
	\end{equation}
	where 
	\begin{equation}
	\bar{z}_k(\bm{\varpi};\mathcal{H}) = \left\{ 
	\begin{array}{l}
\zeta,\;	\textrm{if} \; \vartheta {z}_k(\bm{\varpi};\mathcal{H}) \geq \zeta,\\
-\zeta,\;  \textrm{if} \;  \vartheta {z}_k(\bm{\varpi};\mathcal{H}) \leq -\zeta,\\
\vartheta {z}_k(\bm{\varpi};\mathcal{H}),\;\textrm{otherwise},
	\end{array}
	\right.
	\end{equation}
	and $\zeta>0$ is a constant that is properly chosen according to the smooth parameter $\vartheta$.
	Equivalently, this modification can be viewed as introducing a new piecewise smooth approximation function that transforms \eqref{approximate_step_function} into a linear function when the absolute value of its input is larger than a certain threshold, i.e., 
	\begin{equation} \label{new_approximate_step_function}
	\hat{u}_{\vartheta}(x) =\left\{ \begin{array}{l}  
	\frac{\vartheta e^{-\zeta}}{\left(1+e^{-\zeta}\right)^2} x, \;\textrm{if} \; \vartheta x \geq \zeta,\\
	\frac{\vartheta e^{\zeta}}{\left(1+e^{\zeta}\right)^2} x, \;\textrm{if} \; \vartheta x  \leq -\zeta,\\
     \frac{1}{1+e^{-\vartheta x}},\;\textrm{otherwise}.
	\end{array}
	\right.
	\end{equation}
	In our simulations, utilizing \eqref{new_approximate_step_function} can effectively resolve the ``vanishing gradient'' problem in the proposed Algorithm \ref{algorithm_CSSCA} and accelerate its convergence.
}
\end{remark}
\begin{remark}
\emph{Note that the proposed two-stage CSSCA algorithm (i.e., Algorithm \ref{algorithm_CSSCA}) can be applied to solve the single-user problem \eqref{single_user_problem} as well and it achieves a similar performance as that of the WSMax algorithm in our simulations. However, the WSMax algorithm is more suitable for the single-user case since it is simpler to implement (as it does not require off-the-shelf solvers) and invokes only efficient variable updating steps, which either admit closed-form solutions or can be carried out via simple iterative procedures.}
\end{remark}

\subsection{Complexity Analysis}
The complexity of the first-stage CSSCA algorithm is mainly due to updating the surrogate functions in \eqref{structured_s_function_simplified} (i.e., \textbf{Step 3} in Algorithm \ref{algorithm_CSSCA}) and solving problems \eqref{outage_sub1} or \eqref{outage_sub2} (i.e., \textbf{Step 4} in Algorithm \ref{algorithm_CSSCA}). Specifically, the complexity of updating the surrogate functions in each iteration is dominated by a number of matrix multiplications, which is given by $\mathcal{O}(K^2(L+T_H)((N+M)(KM+N)+NM))$. Besides, since problem \eqref{outage_sub2} contains more optimization variables than problem \eqref{outage_sub1}, the worst-case complexity of \textbf{Step 4} in Algorithm \ref{algorithm_CSSCA} is $\mathcal{O}( K^{0.5}(N+KM+K)^{3})$. Therefore, the complexity of the first-stage CSSCA algorithm is shown to be $\mathcal{C}_1 = \mathcal{O}(I_1(K^2(L+T_H)((N+M)(KM+N)+NM) +  K^{0.5}(N+KM+K)^{3} ))$, where $I_1$ denotes the iteration number. Similarly, the worst-case complexity of the second stage is shown to be $\mathcal{C}_2 =  \mathcal{O}(I_2 (K(L+T_H)M + K^{0.5}(KM+K)^{3})$ with $I_2$ denoting the iteration number. Therefore, the overall complexity can be expressed as $\mathcal{C}_1 + \mathcal{C}_2$.

\section{Simulation Results} \label{section_simulation}
\begin{figure}[!hhh]
	\setlength{\abovecaptionskip}{-0.1cm} 
	\setlength{\belowcaptionskip}{-0.1cm} 
	\centering
	\scalebox{0.45}{\includegraphics{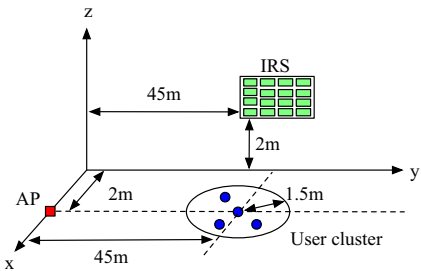}}
	\caption{Simulation setup of the considered IRS-aided MISO downlink system.} 
	\label{user_setup}
\end{figure}

In this section, we provide numerical results to evaluate the performance of the proposed algorithms and draw useful insights. In our simulations, the distance-dependent path loss is modeled as $L = C_0\left({d_{\textrm{link}}}/{D_0}\right)^{-\alpha}$, where $C_0$ is the path loss at the reference distance $D_0 = 1$ meter (m), $d_{\textrm{link}}$ represents the individual link distance and $\alpha$ denotes the path-loss exponent. The path-loss exponents of the AP-user, AP-IRS and IRS-user links are set to $\alpha_{Au} = 3.6$, $\alpha_{AI} = 2.2$ and $\alpha_{Iu} = 2.2$, respectively. A three-dimensional coordinate system is considered where the AP (equipped with a uniform linear array (ULA)) and the IRS (equipped with a uniform rectangular array (UPA)) are located on the $x$-axis and $y$-$z$ plane, respectively. We set $N = N_yN_z$ where $N_y$ and $N_z$ denote the numbers of reflecting elements along the $y$-axis and $z$-axis, respectively. For the purpose of exposition, we fix $N_y = 4$. The reference antenna and reference reflecting element at the AP and IRS are located at $(2\,\textrm{m}, 0, 0)$ and $(0, 45\,\textrm{m}, 2\,\textrm{m})$, respectively, and the users are randomly located on the $x-y$ plane and in a cluster $2$ m away from the IRS with a radius of $1.5$ m, as shown in Fig. \ref{user_setup}. 
To account for small-scale fading, we assume the Rician fading channel model for all channels involved in general. Thus, the AP-IRS channel $\mathbf{G}$ is given by 
\begin{equation}
\mathbf{G} = \sqrt{\frac{\beta_{AI}}{(1+\beta_{AI})}} \mathbf{G}^{\textrm{LoS}} + \sqrt{\frac{1}{(1+\beta_{AI})}} \mathbf{G}^{\textrm{NLoS}},
\end{equation} 
where $\beta_{AI}$ is the Rician factor, and $ \mathbf{G}^{\textrm{LoS}}$ and $ \mathbf{G}^{\textrm{NLoS}}$ represent the deterministic line-of-sight (LoS) and Rayleigh fading non-LoS (NLoS) components, respectively. The AP-user and IRS-user channels are also generated by following the similar procedure and the Rician factors of these two links are denoted by $\beta_{Au}$ and $\beta_{Iu}$, respectively. Other system parameters are set as follows unless otherwise specified: $\sigma_k^2 = -80$ dBm $C_0 = -30$ dB, $N=40$, $M=4$, $N_r = N+1$, $Q = 1$, $\beta_{Au} = \beta_{Iu}   = 0$, $ \beta_{AI} = 3$ dB, $\eta_k = \eta =5\, \textrm{dB},\forall k$, $\epsilon_k = \epsilon = 0.1,\forall k$, $p_{u,k} =p_u = 6 \, \textrm{dBm},\forall k$, $\omega_l=-40$, $\omega_u=10$ and $\Delta \omega = 1$, $\rho^t = (1+t)^{-0.5}$, $\gamma^t = (1+t)^{-0.6}$, $\vartheta = 100$, $\zeta = 8$, $L=10^5$ and $T_H=200$. Note that the specific coefficients in $\{\rho^t, \gamma^t\}$ such as $0.5$ and $0.6$ are tuned to achieve a good empirical convergence speed.

\subsection{Single-User Case}
\subsubsection{Performance Comparison with Benchmark Algorithms}
\begin{figure}[!hhh] 
	\setlength{\abovecaptionskip}{-0.1cm} 
	\setlength{\belowcaptionskip}{-0.1cm} 
	\centering
	\scalebox{0.42}{\includegraphics{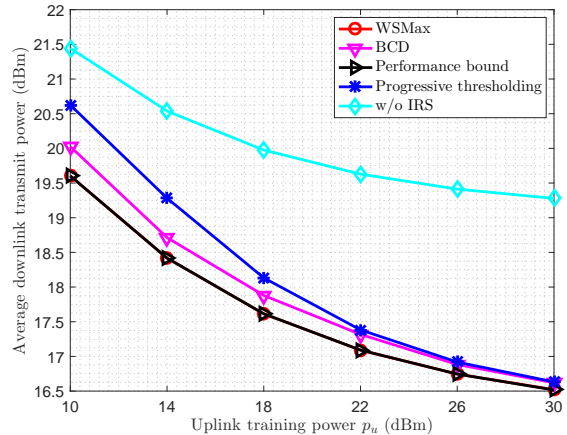}}
	\caption{Performance comparison with benchmark algorithms under different values of $p_u$.}
	\label{fig:SU_compare_bound}
\end{figure}

We first investigate the performance of the proposed WSMax algorithm with fixed $N=10$ and $\eta = 15$ dB, as shown in Fig. \ref{fig:SU_compare_bound}. For comparison, we consider four benchmark algorithms: 1) the performance bound obtained by exhaustively searching over all combinations of the IRS phase shifts and then choose the best one that achieves the minimum transmit power; 2) the BCD algorithm, where the AP transmit power $p$ is found via bisection search and with any given $p$, each IRS phase shift is successively optimized with others fixed until convergence; 3) the conventional scheme by using the MRT beamforming at the AP, but without the IRS; and 4) the progressive thresholding algorithm, where the following SNR-constrained power minimization problem:
  \begin{equation} \label{SNR_constrained_problem}
\begin{aligned}
\min\limits_{\mathbf{w},\; \mathbf{v}} \; & \| \mathbf{w}\|^2\\
\textrm{s.t.}\; & { |( \mathbf{v}^H \hat{\mathbf{H}} + \hat{\mathbf{h}}_{d}^H) \mathbf{w}|^2 }{ }\geq \eta \sigma^2,\;
 v_n \in \mathcal{F}_d,\;\forall n \in \mathcal{N},
\end{aligned}
\end{equation}
 is solved many times using the algorithm in \cite{zhao2019intelligent} and each time with an increased SNR target $\eta$, i.e., $\eta \leftarrow \eta + \delta_{\eta}$ ($\delta_{\eta}$ is set to $0.01$ dB in our simulations), until the outage probability is below $\epsilon$.
 From Fig. \ref{fig:SU_compare_bound}, it is observed that the performance of all algorithms improves as the uplink training power $p_u$ increases, which is because larger $p_u$ implies that the CSI is more accurate and thus less power for data transmission is needed to meet the outage probability constraint. The required AP transmit powers of all algorithms with IRS are significantly lower than that without IRS, which implies that IRS is practically useful even with coarse and low-cost phase shifters and under imperfect CSI.  
 	Besides, the proposed WSMax algorithm can achieve very closely to the performance bound in this simulation setup and it is better than the BCD and progressive thresholding algorithms, especially when the CSI is less accurate.

\subsubsection{Performance Comparison with Different Values of $\omega$}
\begin{figure}[!hhh] 
	\setlength{\abovecaptionskip}{-0.1cm} 
	\setlength{\belowcaptionskip}{-0.1cm} 
	\centering
	\scalebox{0.42}{\includegraphics{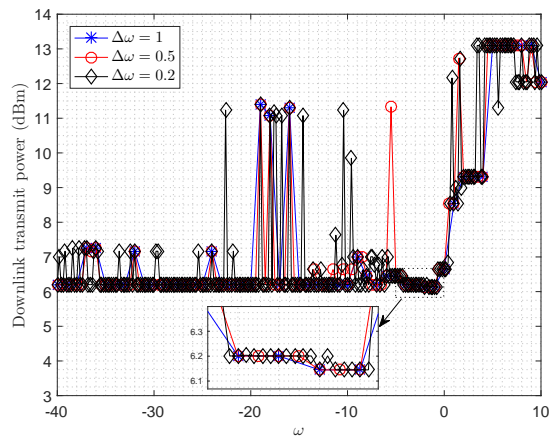}}
	\caption{Downlink transmit power versus weighting factor $\omega$ with different values of $\Delta \omega$.}
	\label{fig:stepsize}
\end{figure}

In Fig. \ref{fig:stepsize}, we plot the downlink transmit power at the AP (for a typical channel realization) versus the weighting factor $\omega$ with different values of the step-size $\Delta \omega$, where $\eta=15$ dB. It can be seen that for many different values of $\omega$ (especially when $\omega <0$), the downlink transmit power is almost constant except for some minor variations and sudden increases. As a result, a very coarse search over $\omega$ is sufficient for the proposed WSMax algorithm to achieve near-optimal performance, which validates the statement in Remark \ref{remark1}.

\subsubsection{Impact of Uplink Training Power, $p_u$}
\begin{figure}[!hhh] 
	\setlength{\abovecaptionskip}{-0.1cm} 
	\setlength{\belowcaptionskip}{-0.1cm} 
	\centering
	\scalebox{0.42}{\includegraphics{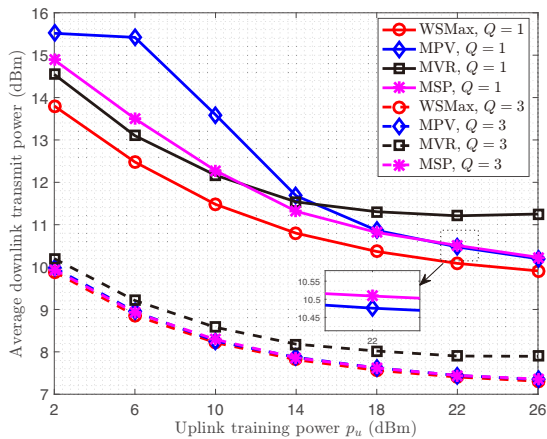}}
	\caption{Average downlink transmit power versus uplink training power, $p_u$.}
	\label{fig:Fig_SU_compare_pu}
\end{figure}

In Fig. \ref{fig:Fig_SU_compare_pu}, we plot the average downlink transmit power at the AP under different values of $p_u$ and $Q$, and provide performance comparison between the proposed WSMax algorithm and the baseline algorithms introduced in Section \ref{Section_heuristic}, where $\eta=15$ dB. First, it is observed that the proposed WSMax algorithm achieves the best performance among the considered counterparts. When $Q=1$, its performance gains over the MPV and MSP maximization algorithms gradually decrease with the increasing of $p_u$,  while the performance gain over the MVR maximization algorithm increases with $p_u$. Besides, maximizing the MVR is better than maximizing the MPV or MSP in the low-$p_u$ regime, which implies that minimizing the variance (corresponding to the negative-$\omega$ case) is more beneficial for minimizing the outage probability when the estimated CSI is less accurate. On the contrary, in the high-$p_u$ regime, maximizing the MPV/MSP is better since the CSI errors are relatively small and it becomes more beneficial to maximize the MSP. Second, we observe that when $Q=3$, the performance gain of the proposed WSMax algorithm is less significant as compared to the case with $Q=1$. This is because as $Q$ increases, the reflection patterns in $\mathbf{V}$ during channel training become near-orthogonal and the CSI errors are less correlated, thus the impact of the variance $\tilde{\mathbf{v}}^H \bar{\mathbf{V}} \tilde{\mathbf{v}}$ is minor. As a result, maximizing the MPV and MSP is better than maximizing the MVR in this case.

\subsubsection{Impact of SNR Target, $\eta$}
\begin{figure}[!hhh] 
	\setlength{\abovecaptionskip}{-0.1cm} 
	\setlength{\belowcaptionskip}{-0.1cm} 
	\centering
	\scalebox{0.42}{\includegraphics{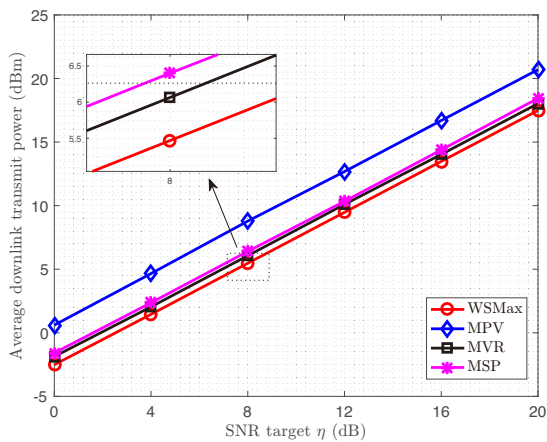}}
	\caption{Average downlink transmit power versus SNR target, $\eta$.}
	\label{fig:Fig_SU_compare_eta}
\end{figure}

In Fig. \ref{fig:Fig_SU_compare_eta}, we investigate the average downlink transmit power at the AP versus the SNR target $\eta$. It can be seen that the required power of all the considered algorithms increases with $\eta$, which is reasonable since more power is needed to achieve higher SNR. In addition, we observe that the performance gains of the proposed WSMax algorithm over the baseline algorithms are almost invariant with different values of $\eta$. This is due to the fact that $\eta$ scales proportionally with $p$ and thus does not affect the optimization of the IRS phase shifts, as can be seen from \eqref{v_subproblem_eq2}.

\subsubsection{Impact of Outage Probability Target, $\epsilon$}
\begin{figure}[!hhh] 
	\setlength{\abovecaptionskip}{-0.1cm} 
	\setlength{\belowcaptionskip}{-0.1cm} 
	\centering
	\scalebox{0.42}{\includegraphics{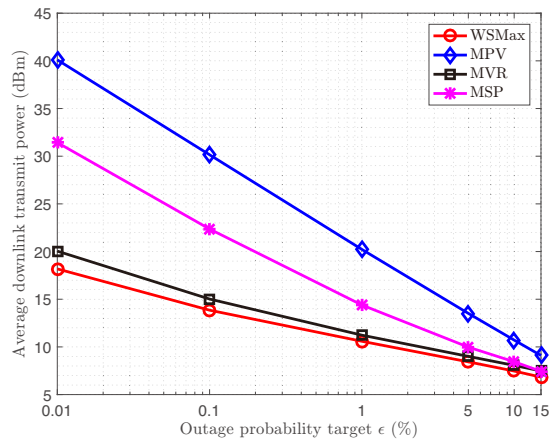}}
	\caption{Average downlink transmit power versus outage probability target, $\epsilon$.}
	\label{fig:Fig_SU_compare_epsilon}
\end{figure}

Next, in Fig. \ref{fig:Fig_SU_compare_epsilon}, we plot the average downlink transmit power at the AP versus the outage probability target $\epsilon$, where $\eta = 10$ dB. First, it is observed that the downlink transmit power of all algorithms decreases as $\epsilon$ increases, which shows that less power is needed if the outage probability requirement is less stringent. Second, we observe that the proposed WSMax algorithm achieves the lowest transmit power and its performance gains over the baseline algorithms enlarges with the decreasing of $\epsilon$. Therefore, the proposed WSMax algorithm is able to significantly reduce the required transmit power at the AP, especially for more reliable transmissions (corresponding to smaller values of $\epsilon$).

\subsubsection{Impact of Number of Reflecting Elements, $N$}
\begin{figure}[t] 
	\setlength{\abovecaptionskip}{-0.1cm} 
	\setlength{\belowcaptionskip}{-0.1cm} 
	\centering
	\scalebox{0.42}{\includegraphics{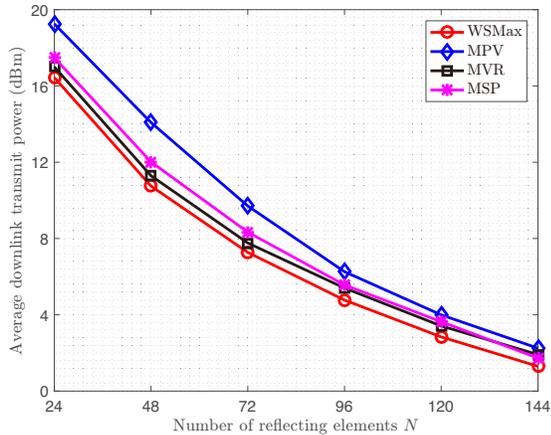}}
	\caption{Average downlink transmit power versus number of IRS elements, $N$.}
	\label{fig:Fig_SU_compare_N}
\end{figure}

In Fig. \ref{fig:Fig_SU_compare_N}, we plot the average downlink transmit power at the AP versus the number of IRS reflecting elements $N$, where $\eta = 15$ dB. It is observed that  the downlink transmit power of all algorithms decreases when $N$ increases, which is reasonable since larger $N$ leads to higher aperture gain and offers more flexibility when designing the passive beamforming with discrete phase shifts at the IRS. Besides, we can see that the performance gain of the proposed WSMax algorithm over the MVR maximization algorithm slightly increases with $N$ and that over the MPV maximization algorithm decreases with $N$. This is because when $N$ increases, the IRS becomes more effective in maximizing the MSP and the impact of the variance is less significant; therefore, it becomes less useful to minimize the variance through maximizing the MVR. This also explains why the performance of the MSP maximization algorithm is better than that of the MVR maximization algorithm in the large-$N$ regime. Moreover, the performance gain of the proposed WSMax algorithm over the MSP maximization algorithm is almost invariant with different values of $N$, which is because the latter ignores the optimization of the variance. 

\subsection{Multiuser Case}
\begin{figure}[!hhh] 
	\setlength{\abovecaptionskip}{-0.1cm} 
	\setlength{\belowcaptionskip}{-0.1cm} 
	\centering
	\scalebox{0.34}{\includegraphics{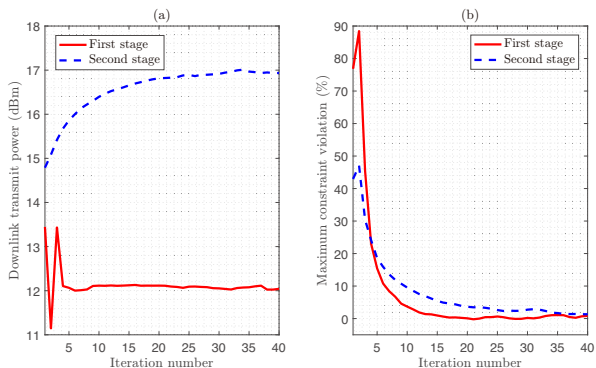}}
	\caption{Convergence behavior of the proposed two-stage CSSCA algorithm.}
	\label{fig:Fig_MU_convergence}
\end{figure}

In this subsection, we consider the multiuser system with $K \geq 2$ users and the AP is equipped with $M = 6$ antennas, with $p_u = 18$ dBm. 
We first illustrate in Fig. \ref{fig:Fig_MU_convergence} the convergence behavior of the proposed two-stage CSSCA algorithm by plotting the required downlink transmit power and maximum constraint violation (i.e., the highest outage probability among the users $\max_k\{f_k^t\}$ minus the outage probability target $\epsilon$) versus the number of iterations with $K=4$. From Fig. \ref{fig:Fig_MU_convergence}, we can observe that although the curves are not necessarily monotonic due to the stochastic nature of the proposed algorithm, it is able to converge in about $40$ iterations (for both stages) and the maximum constraint violation $\max_k\{f_k^t\} - \epsilon$ gradually converges to zero as the iteration number increases.

\begin{figure}[!hhh] 
	\setlength{\abovecaptionskip}{-0.1cm} 
	\setlength{\belowcaptionskip}{-0.1cm} 
	\centering
	\scalebox{0.42}{\includegraphics{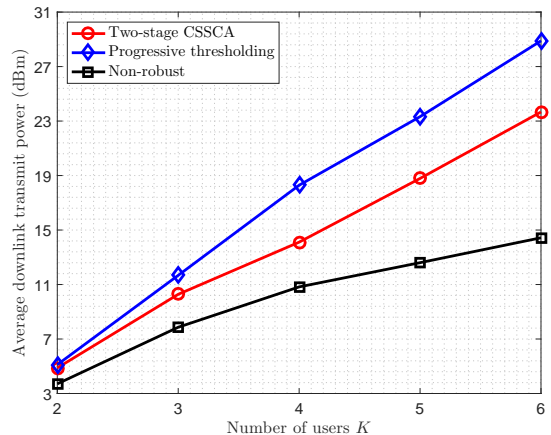}}
	\caption{Average downlink transmit power versus number of users, $K$}
	\label{fig:Fig_MU_compare_K}
\end{figure}

Finally, in Fig. \ref{fig:Fig_MU_compare_K}, we investigate the average downlink transmit power at the AP versus the number of users, $K$. Similar to that in Fig. \ref{fig:SU_compare_bound}, the performance of the progressive thresholding algorithm is provided for comparison, where the underlying SINR-constrained power minimization problem is solved by using the algorithm in \cite{Wu2019Discrete}. The non-robust scheme is obtained by designing $\{\mathbf{w}_k\}$ and $\mathbf{v}$ based on the estimated CSI and ignoring the outage probability constraints (thus cannot guarantee any outage performance). It is observed that the proposed two-stage CSSCA algorithm outperforms the progressive thresholding algorithm and the performance gain is more pronounced when $K$ increases. This is because the multiuser interference due to imperfect CSI is more severe with larger $K$ and for the progressive thresholding algorithm, it becomes more difficult to design $\{\mathbf{w}_k\}$ and $\mathbf{v}$ to guarantee outage probability by simply increasing $\eta$. For the same reason, the performance gap between the proposed algorithm and the non-robust scheme enlarges with the increasing of $K$, which implies that more power is needed as a price paid for guaranteed outage performance.

\section{Conclusion} \label{Section_conclusion}
In this paper, we studied an outage-constrained power minimization problem for joint active and passive beamforming design in an IRS-aided communication system, under correlated CSI errors. We proposed two efficient algorithms, i.e., the WSMax algorithm and two-stage CSSCA algorithm, for the single-user and multiuser cases, respectively. Simulation results showed that the proposed algorithms can effectively reduce the transmit power at the AP with guaranteed outage performance, especially when the channel training resources are limited.

\bibliographystyle{IEEETran}
\bibliography{references}

\end{document}